\documentclass[sigconf,review=false,natbib=true,anonymous=false]{acmart}

\copyrightyear{2024}
\acmYear{2024}
\setcopyright{rightsretained}
\acmConference[SIGIR '24]{Proceedings of the 47th International ACM SIGIR Conference on Research and Development in Information Retrieval}{July 14--18, 2024}{Washington, DC, USA}
\acmBooktitle{Proceedings of the 47th International ACM SIGIR Conference on Research and Development in Information Retrieval (SIGIR '24), July 14--18, 2024, Washington, DC, USA}
\acmDOI{10.1145/3626772.3657869}
\acmISBN{979-8-4007-0431-4/24/07}

\usepackage{xspace}
\usepackage{bbm}
\usepackage{enumitem}

\begin{document}
\title{ProCIS: A Benchmark for Proactive Retrieval in Conversations}

\author{Chris Samarinas}
\affiliation{%
  \institution{University of Massachusetts Amherst}
  \city{Amherst}
  \state{MA}
  \country{United States}
}
\email{csamarinas@cs.umass.edu}

\author{Hamed Zamani}
\affiliation{%
  \institution{University of Massachusetts Amherst}
  \city{Amherst}
  \state{MA}
  \country{United States}
}
\email{zamani@cs.umass.edu}


\begin{abstract}
The field of conversational information seeking, which is rapidly gaining interest in both academia and industry, is changing how we interact with search engines through natural language interactions. Existing datasets and methods are mostly evaluating \emph{reactive} conversational information seeking systems that solely provide response to every query from the user. We identify a gap in building and evaluating \emph{proactive} conversational information seeking systems that can monitor a multi-party human conversation and proactively engage in the conversation at an opportune moment by retrieving useful resources and suggestions. In this paper, we introduce a large-scale dataset for proactive document retrieval that consists of over 2.8 million conversations. We conduct crowdsourcing experiments to obtain high-quality and relatively complete relevance judgments through depth-k pooling. We also collect annotations related to the parts of the conversation that are related to each document, enabling us to evaluate proactive retrieval systems. We introduce normalized proactive discounted cumulative gain (npDCG) for evaluating these systems, and further provide benchmark results for a wide range of models, including a novel model we developed for this task. We believe that the developed dataset, called ProCIS, paves the path towards developing proactive conversational information seeking systems.


\end{abstract}



\maketitle

\section{Introduction}
Conversational information seeking (CIS) has been identified as an emerging subfield of research with information retrieval and related disciplines \cite{zamani2023conversational}. CIS systems are revolutionizing the way users seek and access information through natural language dialogues \cite{Amherst2023, samarinas2024simulating}. The advent of large language models (LLMs), such as LLaMA \cite{touvron2023llama}, Mistral \cite{jiang2023mistral}, and GPT-4 \cite{openai2023gpt4}, in the past year has opened up new opportunities for improving these systems. 

Existing models for CIS are generally designed based on a \textit{``query-response''} paradigm, where the user starts the interaction by submitting a search query and the system responds with a search result (e.g., a ranked-list of passages or documents that may include snippets, a direct natural language answer to the submitted query, one or more entity cards, or a combination of them). In this paradigm, users can interact with the search results and/or try a different query which will be answered by another search result. This process repeats until the user terminates the search session. Popular conversational question answering benchmarks, such as QuAC \cite{choi-etal-2018-quac} and CoQA \cite{reddy-etal-2019-coqa}, and TREC Conversational Assistance Track (CAsT) \cite{owoicho2022trec} have been mainly focusing on evaluating CIS systems based on such paradigm. However, this is not an optimal interaction design for CIR systems and they must be able to perform \emph{mixed-initiative interactions} \cite{10.1145/3459637.3482231}. Asking clarifying questions \cite{Zamani2020ClarAnalysis,Zamani2020ClarGeneration,Zamani2020MIMICS,Aliannejadi2019Qulac,Braslavski:2017,Rao:2018,Stoyanchev:2014,10.1145/3488560.3498440} is a type of mixed-initiative interaction that has been extensively studied in the context of CIS. For more information about mixed-initiative CIS, refer to \cite[Chapter 6]{zamani2023conversational}.

This paper focuses on \emph{proactive interactions} as another type of mixed-initiative CIS \cite{related13}. Proactive CIS, despite numerous applications, has been relatively under-explored. A reason for this is lack of data for training and evaluating proactive systems. This paper attempts to bridge this gap and presents a new large-scale data collection for proactive CIS. To this aim, we focus on \emph{document retrieval as proactive contextual suggestion to multi-party human conversations}.

Imagine multiple users are interacting with each other and a CIS agent is monitoring this conversation. This can be a conversation in a forum, in a group chat such as Slack channel,\footnote{\url{https://slack.com/}} or a bot listening to an ongoing conversation between people in a meeting room.\footnote{In some of these scenarios, there are privacy considerations that should be taken into account. These issues are outside the scope of this work, yet important regardless.}
While users are conversing, an agent may engage in the conversation by providing a useful suggestion or by verifying the factually of the claims in the conversation. Such proactive informational interactions are of interest to this work. Therefore, we introduce the task of proactive document retrieval that given a sequence of utterances decides whether to engage in the conversation with a retrieval result list or not.

To build a large-scale data collection for this work, we use a Wikipedia dump with about 5 million English articles as a knowledge source, i.e., retrieval collection, that a CIS agent can use to proactive engage in a conversation. We then collected Reddit threads\footnote{We obtain Reddit threads from \url{https://github.com/ArthurHeitmann/arctic_shift}. Any use of the data must be in accordance with the Reddit's term of services.} in which multiple users are interacting with each other about a topic and there exists a link to Wikipedia articles in that thread. We assume that such threads are likely to be informational and we can build systems that find relevant Wikipedia pages once they are excluded from the conversation. After careful filtering and data cleaning, we end up with a large-scale dataset with over 2.8 million conversations, called \emph{ProCIS}. We provide multiple data splits for training, development, and testing. To have a reliable test set, we conduct crowdsourcing experiments to annotate documents for each conversation. We not only collect relevance judgment, but also ask annotators to highlight the parts of conversation that are related to each relevant document. This enables us to evaluate proactive systems by knowing what documents provide useful suggestions to each conversation utterance. The crowdsourcing experiments are conducted using Amazon Mechanical Turk and the depth-$k$ pooling approach is applied to construct the document pool for annotation.

We further provide an evaluation methodology for evaluating proactive document retrieval systems. We introduce normalized proactive discounted cumulative gain ($npDCG$) for evaluating such methods. We also adopt the developed dataset for \emph{reactive} document retrieval methods for contextual suggestion. There can also be real-world applications for such reactive scenarios, where users explicitly ask a bot to engage in a multi-party human conversation and provide useful suggestions. We evaluate a term-matching retrieval model, a neural sparse retrieval model, a single-vector dense retrieval model, and a multi-vector dense retrieval model on the evaluated benchmarks. We also introduce a novel approach that is employing LLMs for query generation and result filtering. 

We believe the benchmark results presented in this paper smooth the path towards developing advanced proactive CIS systems. We release the data, code, and benchmark results for research purposes: \url{https://github.com/algoprog/ProCIS}.

\section{Related Work}

This section reviews the literature on the three main areas related to our work.

\subsection{Proactive Search Systems}

Proactive search systems, designed to anticipate user needs and provide relevant information without explicit queries, are gaining attention due to their potential to enhance user experience and improve search efficiency. A study on the effectiveness of proactive search systems proposed a framework to evaluate such systems based on the correlation between the expected and predicted outcomes \cite{rel3}. This study demonstrated the potential of proactive search systems to recommend documents that could help users accomplish their tasks without explicit queries. 

Another work introduced the concept of \emph{information fostering}, a proactive approach that predicts potential issues and provides help to overcome them \cite{rel4}. This approach goes beyond recommending queries and documents and suggests strategies and people, thereby offering a more comprehensive support system for information seekers. 
The use of Wikipedia concepts in proactive information retrieval was explored in a study on improving retrieval on noisy text \cite{rel5}. The study demonstrated the potential of Wikipedia concepts to provide relevance signals and improve precision in proactive information retrieval. 

A study on proactive search support in conversations demonstrated how a proactive search agent could augment conversations by retrieving and presenting information related to the conversation \cite{rel1}. The study highlighted the potential of proactive search systems to affect the topical structure of conversations and reduce the need for explicit search activity. Recently, \citet{related13} highlighted the opportunities and challenges in proactive interactions in conversational information seeking. 

This work complements the literature on proactive search by building a large-scale dataset, introducing an evaluation methodology, and presenting benchmark results. For more information about proactive search systems, refer to the tutorial recently presented by \citet{10.1145/3539597.3572724}.

\begin{table*}[t]
\small
\caption{Statistics of each data split in the ProCIS dataset.}
\begin{tabular}{lccccp{0.2\linewidth}}
\toprule
 &
\textbf{train} &
\textbf{dev} &
\textbf{future-dev} &
\textbf{test} \\ \midrule
Total conversations &
  2,830,107 &
  4165 &
  3385 & 
  100 \\
Total posts                                  & 1,893,201 & 4165     & 3385 & 100 \\ 
Number of subreddits covered                 & 34,785    & 1563    & 1659 & 100  \\
Total unique users in the conversations & 2,284,841 & 10,896 & 7,920 & 309 \\ \midrule
Average number of turns                      & 5.41 ($\pm$ 7.81)    & 4.91 ($\pm$ 3.60)   & 4.48 ($\pm$ 3.30) & 4.49 ($\pm$ 1.60) \\
Average number of words per conversation           & 406.01 ($\pm$ 774.67)  & 359.19 ($\pm$ 734.95) & 325.36 ($\pm$ 609.58) & 173.85 ($\pm$ 101.22) \\
Average number of words per turn             & 70.54 ($\pm$ 82.38)   & 68.77 ($\pm$ 74.80)  & 72.55 ($\pm$ 85.37) & 41.58 ($\pm$ 26.49) \\
Average number of Wikipedia links per conversation & 1.71 ($\pm$ 2.46)    & 1.90 ($\pm$ 3.03)   & 1.15 ($\pm$ 0.57) & 1.15 ($\pm$ 0.46) \\
Average number of unique users per conversation & 3.17 ($\pm$ 1.41) & 2.93 ($\pm$ 1.16) & 2.88 ($\pm$ 1.11) & 3.41 ($\pm$ 1.39) \\
Average number of comments per user & 6.71 ($\pm$ 462.74) & 1.88 ($\pm$ 8.21) & 1.92 ($\pm$ 12.93) & 1.45 ($\pm$ 2.49) \\
\bottomrule
\end{tabular}
\label{dataset-stats}
\end{table*}

\subsection{Conversational Information Seeking}

CIS is a rapidly evolving field that mainly focuses on retrieving information in the context of conversations \cite{related11}. The Conversational Assistance Track (CAsT) \cite{rel7} was a significant initiative in this direction, aiming to facilitate Conversational Information Seeking (CIS) research and create a large-scale reusable test collection for conversational search systems. 


Several studies have proposed models and theories to address the challenges unique to CIS. For instance, one study proposed a theoretical model of a conversational system that implements a small set of properties derived from past work on human conversations \cite{related1}. Another study proposed a Conversational Dense Retrieval model that learns contextualized embeddings for multi-turn conversational queries \cite{related3}. 


The role of retrieval in CIS has also been explored. One study introduced an open-retrieval conversational question answering task, where evidence is retrieved from a large collection before extracting answers \cite{related5}. Another study proposed a pipeline for passage retrieval in a conversational search setting, comprising conversational term selection and multi-view reranking \cite{related7}. 

Datasets play a crucial role in the development and evaluation of conversational search models. Several studies have introduced new datasets to facilitate research in this area. For instance, one study introduced MANtIS, a large-scale dataset containing multi-domain and grounded information seeking dialogues \cite{related4}. Another study created a dataset, OR-QuAC, to facilitate research on open-retrieval conversational question answering \cite{related5}. The dataset created by \citet{10.1145/3578337.3605139} is perhaps the most similar data to ProCIS. It is also based on Reddit thread, however, ProCIS has multiple advantages in comparison; it is more than an order of magnitude larger in terms of both training examples and corpus size, it has a carefully annotated test set and it uses a corpus of clean Wikipedia articles which can be more useful for research experiments that involve specific concepts. In a similar vein, the RCD-2020 track \cite{DBLP:conf/fire/GangulyPVS20} introduced another small-scale dataset. However, the conversations in this dataset are derived from short movie dialogues, resulting in brief and simplistic exchanges. Consequently, the dataset is primarily suitable for limited evaluation purposes rather than comprehensive conversational modeling.

CIS is a complex and challenging task that requires comprehensive understanding of the conversational inputs, effective query reformulation, and efficient retrieval methods. The studies discussed in this section have made significant contributions towards addressing these challenges. However, there is still much work to be done, particularly in the area of proactive retrieval in conversations, which is the focus of our work.

\subsection{Information Filtering}
Information filtering is a closely related area to proactive conversational information seeking. The TREC Filtering Track \cite{trecfiltering} and CLEF INFILE \cite{clefinfile} focused on the multiple filtering tasks including adaptive filtering, where systems aim to select relevant documents from a stream of incoming documents based on a user's profile. These tracks have contributed to the development of effective filtering techniques based on threshold optimization \cite{arampatzis2009threshold} and profile adaptation \cite{gauch2007profile,contentbasedrecsys}.

\begin{figure*}[t]
  \centering
  \includegraphics[width=0.7\textwidth]{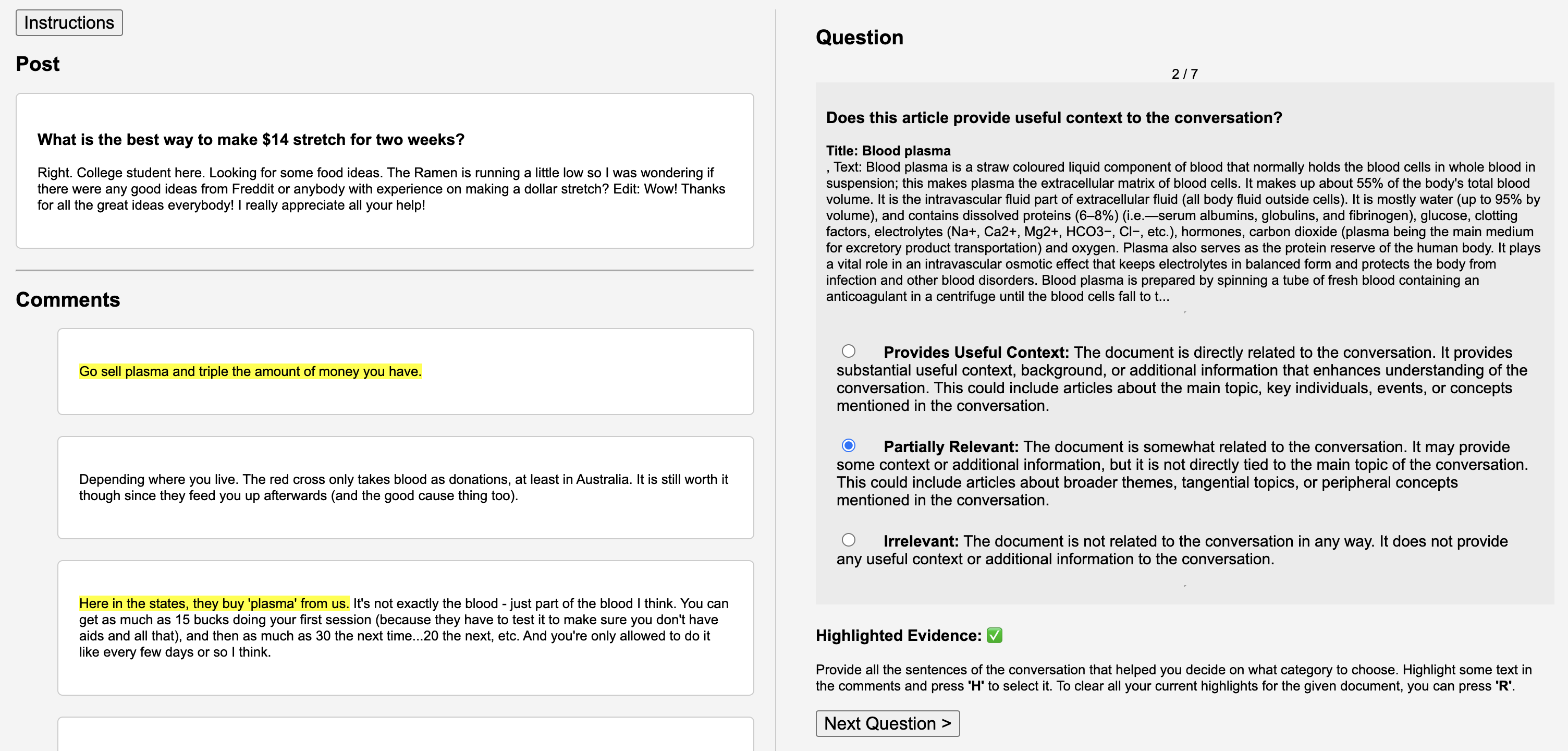}
  \caption{The online annotation interface for the crowd-sourcing of the test set.}
  \label{fig:annotation-ui}
\end{figure*}

\section{Data Collection}
Our goal is to construct a large-scale dataset that enables proactive retrieval research for multi-party human conversations. To this aim, we focus on Wikipedia as the retrieval corpus for providing information proactively and use Reddit threads to obtain multi-party human conversations. In order to focus on the conversation threads that are likely to benefit from information seeking, we only collect conversations that contain one or more hyperlinks to a Wikipedia article. To gather this data, we utilize the publicly available dumps of Reddit posts from pushshift.io,\footnote{Currently available at: \url{https://github.com/ArthurHeitmann/arctic_shift}} collected from 2005 until 2022. We applied several filters to ensure the data quality, such as removing not-suitable-for-work (NSFW) content, posts with external links, or embedded media. We also excluded non-English content and removed HTML formatting and link mentions. We then sampled nested threads where each comment in the chain is the child of the previous comment. 

As the document corpus, we used a pre-processed dump of 5,315,384 Wikipedia articles,\footnote{Available at \url{https://github.com/tscheepers/Wikipedia-Summary-Dataset}} and we mapped all the extracted Wikipedia links in the collected Reddit threads to the articles in this corpus. We further excluded the threads with links to Wikipedia articles that do not exist in the corpus. After applying all the filters, the final dataset comprised 2,830,107 conversations from 1,893,201 unique posts. 


\paragraph{\textbf{Data Splits}} We split the data to four subsets: train, dev, future-dev, and test. The three subsets of train, dev, and test are split randomly, while the future-dev set only contains conversations that follow the conversations in the training set chronologically. This split can be used for evaluating the generalization capabilities of retrieval models in potentially new emerging concepts and topics not seen during training. The test split was sampled from 100 unique random subreddits, all from posts with at least a Reddit score of 20 to ensure high quality. The relevance annotation in train, dev, and future-dev is sparse and sometimes noisy due to topic shifts. We assume a Wikipedia article is relevant to a conversation, if its URL is mentioned in the conversation thread. After some analysis through crowdsourcing, we found that 63\% of the mentioned Wikipedia articles actually provide useful context to the conversation. This enables us to construct a large-scale dataset with sparse noisy annotation. According to lessons learned from the MS MARCO collection \cite{nguyen2016ms}, large-scale sparse annotations can be quite impactful in training retrieval models. To ensure comprehensive evaluation, we conduct pooling and collect human annotations for the documents in the test set. Therefore, even though the test set contains only 100 conversations, it has on average 8.02 relevant documents per conversation, much higher than the 1.15 (potentially irrelevant) links on average that the users mentioned originally on Reddit. Table \ref{dataset-stats} presents the statistics of ProCIS. 


\begin{figure*}[t]
  \centering
  \begin{minipage}{0.32\textwidth}
    \centering
    \includegraphics[width=\textwidth]{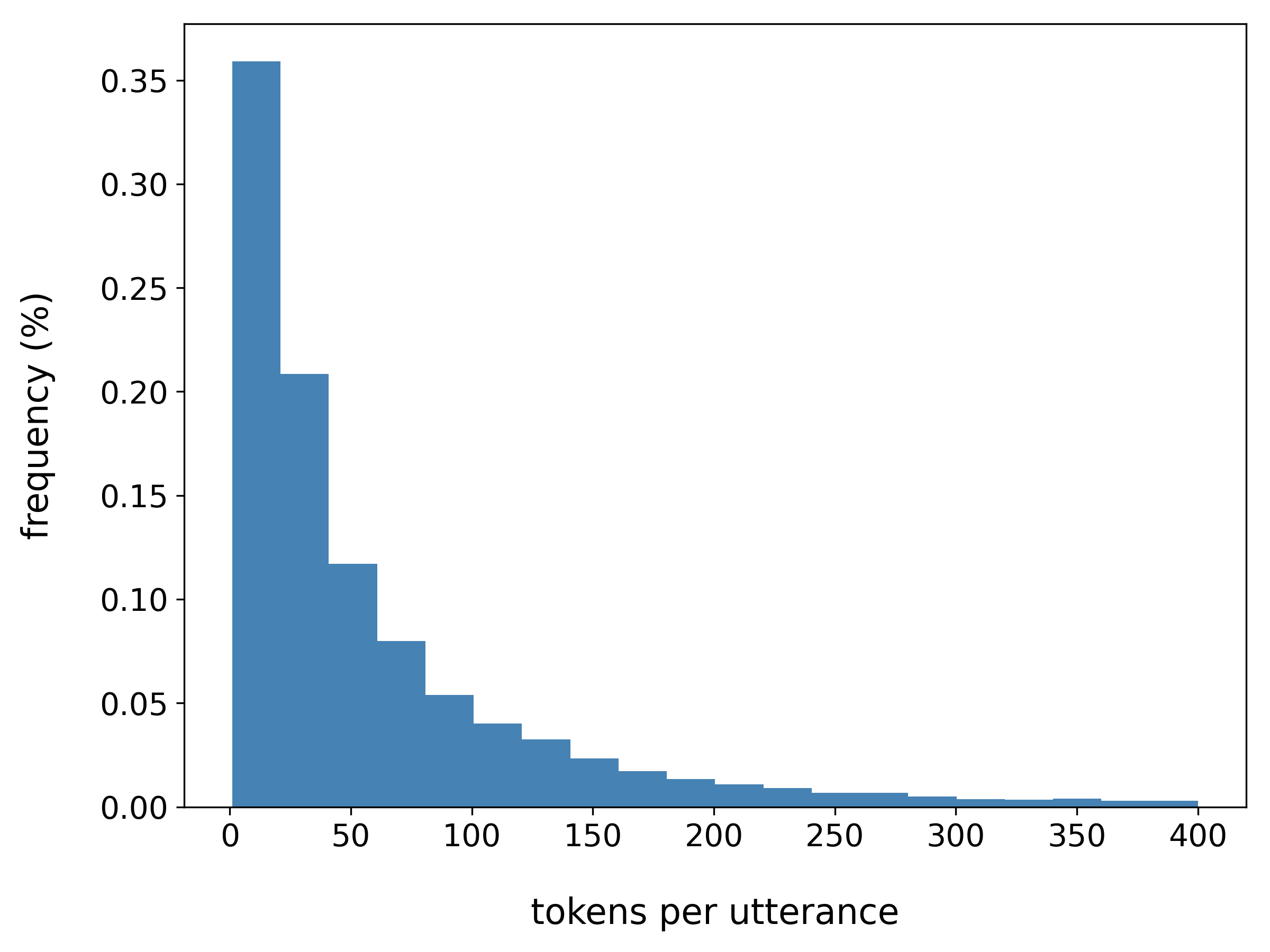}
    \caption{\small Utterance length (\# tokens) distribution in ProCIS.}
    \label{fig:comment-lengths-plot}
  \end{minipage}\hfill
  \begin{minipage}{0.32\textwidth}
    \centering
    \small
    \includegraphics[width=\textwidth]{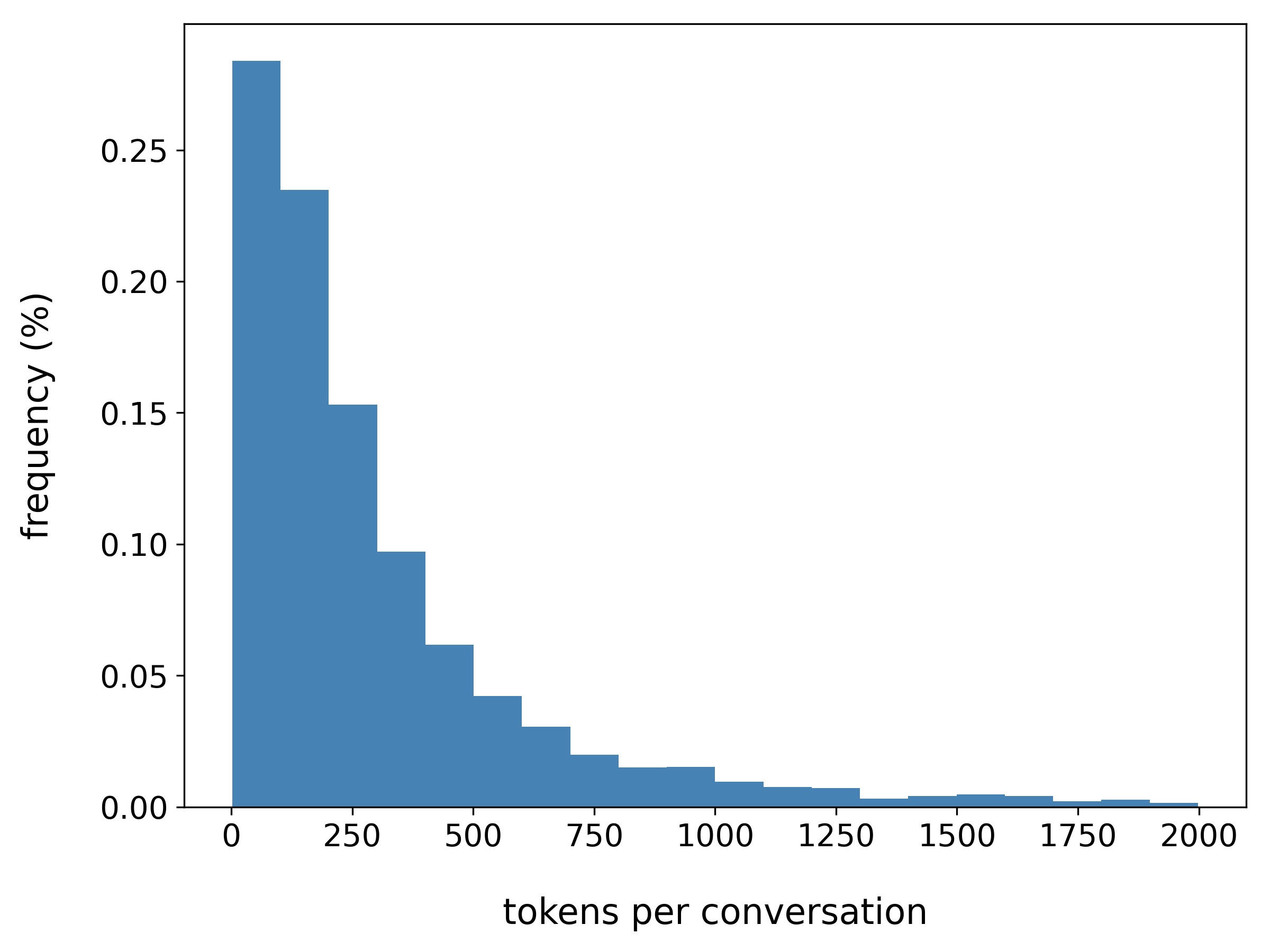}
    \caption{\small Conversation length (\# tokens) distribution in ProCIS.}
    \label{fig:thread-token-lengths-plot}
  \end{minipage}\hfill
  \begin{minipage}{0.32\textwidth}
    \centering
    \small
    \includegraphics[width=\textwidth]{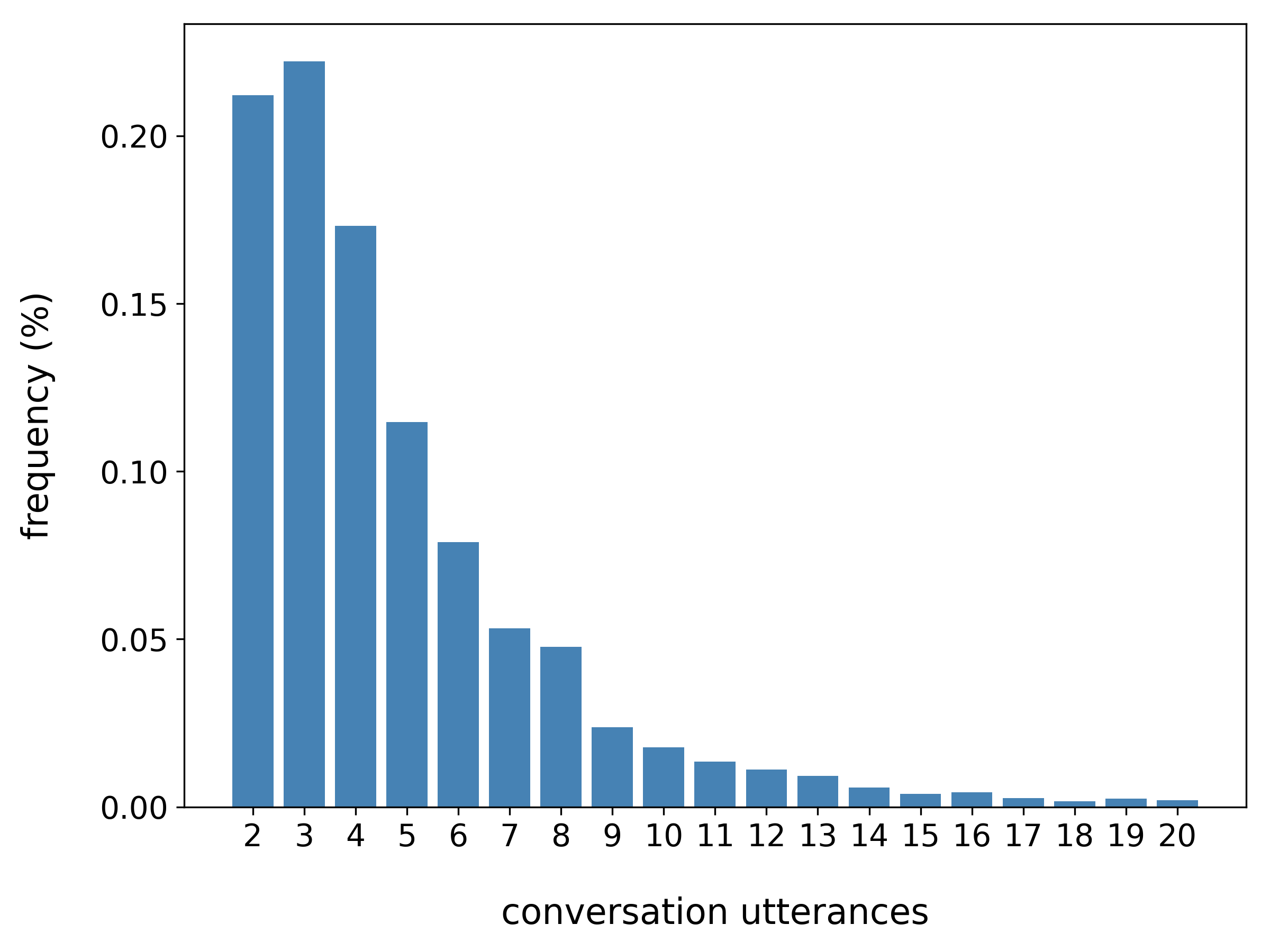}
      \caption{\small Conversation length (\# turns) distribution in ProCIS.}
      \label{fig:turns-plot}
  \end{minipage}
  \vspace{0.1cm}
\end{figure*}


\begin{table*}[ht]
\small
\caption{The 60 most frequent categories (subreddits) in the ProCIS training set.}
\begin{tabular}{lr|lr|lr}
\toprule
\textbf{Subreddit} & \textbf{Frequency} & \textbf{Subreddit} & \textbf{Frequency} & \textbf{Subreddit} & \textbf{Frequency} \\ \midrule
AskReddit & 500,419 (38.19\%) & neoliberal & 12,751 (0.97\%) & UnresolvedMysteries & 7062 (0.54\%) \\ 
askscience & 58,527 (4.47\%) & AskAnAmerican & 12,664 (0.97\%) & exmormon & 7040 (0.54\%) \\ 
explainlikeimfive & 55,063 (4.20\%) & CapitalismVSocialism & 12,529 (0.96\%) & AskMen & 6753 (0.52\%) \\ 
IAmA & 44,315 (3.38\%) & CFB & 12,232 (0.93\%) & OutOfTheLoop & 6745 (0.51\%) \\ 
changemyview & 34,106 (2.60\%) & soccer & 12,179 (0.93\%) & hockey & 6700 (0.51\%) \\ 
atheism & 31,717 (2.42\%) & math & 12,108 (0.92\%) & books & 6636 (0.51\%) \\ 
politics & 28,388 (2.17\%) & bestof & 11,737 (0.90\%) & Jokes & 6546 (0.50\%) \\ 
DebateReligion & 25,419 (1.94\%) & anime & 11,244 (0.86\%) & unitedkingdom & 6164 (0.47\%) \\ 
Christianity & 24,170 (1.84\%) & leagueoflegends & 10,680 (0.82\%) & reddit.com & 6087 (0.46\%) \\ 
PoliticalDiscussion & 22,467 (1.71\%) & europe & 10,650 (0.81\%) & learnprogramming & 5989 (0.46\%) \\ 
Showerthoughts & 20,035 (1.53\%) & Fitness & 9375 (0.72\%) & buildapc & 5945 (0.45\%) \\ 
SubredditDrama & 18,751 (1.43\%) & whowouldwin & 9198 (0.70\%) & cars & 5928 (0.45\%) \\ 
NoStupidQuestions & 18,286 (1.40\%) & DebateAChristian & 9066 (0.69\%) & talesfromtechsupport & 5836 (0.45\%) \\ 
history & 17,200 (1.31\%) & Libertarian & 8889 (0.68\%) & asoiaf & 5804 (0.44\%) \\ 
AskHistorians & 16,709 (1.28\%) & nba & 8883 (0.68\%) & space & 5566 (0.42\%) \\ 
conspiracy & 15,466 (1.18\%) & india & 8628 (0.66\%) & guns & 5445 (0.42\%) \\ 
nfl & 14,688 (1.12\%) & tipofmytongue & 7795 (0.59\%) & NeutralPolitics & 5423 (0.41\%) \\ 
unpopularopinion & 13,828 (1.06\%) & DebateAnAtheist & 7718 (0.59\%) & ukpolitics & 5351 (0.41\%) \\ 
AskEurope & 13,206 (1.01\%) & AskTrumpSupporters & 7564 (0.58\%) & Drugs & 5306 (0.40\%) \\ 
movies & 12,899 (0.98\%) & Bitcoin & 7160 (0.55\%) & LifeProTips & 5277 (0.40\%) \\ 
\bottomrule
\end{tabular} 
\label{subreddits}
\end{table*}

\paragraph{\textbf{Relevance Assessment for the Test Set}}
Reliable evaluation requires complete relevance annotation. In order to collect relevance annotations for the constructed test set, we follow the pooling guideline from TREC. For each conversation in the test split, we created 5 pools of up to 10 document candidates from the following models: BM25, SPLADE, ANCE, ColBERT and LMGR with GPT-4 (refer to Section \ref{sec:method} for LMGR). The reason for choosing these models is to obtain a diverse pool of documents. BM25 is a term-matching model, while SPLADE is a neural sparse retrieval model, ANCE and ColBERT are single- and multi-vector dense retrieval models, and LMGR is an effective generative model for retrieval. The supervised neural models were fine-tuned on the ProCIS training set before producing the pools. 

Once the pools are constructed, we designed a careful crowdsourcing experiment on Amazon's Mechanical Turk for data annotation. Each Human Intelligence Task (HIT) asks the crowdworkers to do the following:
\begin{enumerate}
    \item \textbf{Understand} the Conversation: read the Reddit post title, content, and comments one by one to understand the main topic and subtopics of the conversation. The crowdworker should read each utterance and click on `next' to see the next utterance.
    \item \textbf{Summarize} the whole conversation in 1-2 sentences covering the main topics and themes.
    \item \textbf{Read and annotate} each document by selecting one of the three relevance levels: provides useful context, partially relevant or irrelevant (see below). 
    \item \textbf{Highlight evidence}: In case of a relevant or partially relevant document, highlight all the sentences of the conversation that are related to the document. 
\end{enumerate}

These steps ensure that the crowdworkers spend enough time to thoroughly understand each conversation and help us easily detect and discard low quality annotations. To facilitate the annotation process, we built a custom interface with the following features: it first displays annotation instructions, then expects the worker to read the comments one at a time, expects a brief summary with a minimum required length of six words and then displays documents one by one for annotation. For each document, it asks for a relevance level option, and in case of relevance or partial relevance, the worker is required to highlight some text in the conversation (see Figure~\ref{fig:annotation-ui}). We ask for three-level graded relevance annotation with respect to the following definitions:

\begin{enumerate}
\item \textbf{Provides Useful Context (Label 2)}: The document is directly related to the conversation. It provides substantial useful context, background, or additional information that enhances understanding of the conversation. This could include articles about the main topic, key individuals, events, or concepts mentioned in the conversation.

\item \textbf{Partially Relevant (Label 1)}: The document is somewhat related to the conversation. It may provide some context or additional information, but it is not directly tied to the main topic of the conversation. This could include articles about broader themes, tangential topics, or peripheral concepts mentioned in the conversation.

\item \textbf{Irrelevant (Label 0)}: The document is not related to the conversation in any way. It does not provide any useful context or additional information to the conversation.
\end{enumerate}

In each HIT, we display a conversation between 3 and 10 utterances and up to 500 words long. In total we created 1000 HITs in 8 batches and collected 4207 relevance assessments with supporting evidence. Each HIT was completed by 3 different workers. We limited the HITs to adult workers from the US, UK, Australia and Ireland, with over 98\% approval rate who have completed at least 5,000 assignments. The inter-annotator agreement was 0.6482 in Fleiss' $\kappa$ score. For the final labels we used majority voting. If two annotators picked partially relevant and relevant as labels for one document and the third one irrelevant, we assigned relevant as the final label. The ideal positions of the annotated documents were identified based on the earliest position of the highlighted supported evidence from the 3 workers for each HIT. The total annotation cost was \$3500 for \$1.16 per HIT. The annotation interface is released for transparency and future usage. 

\section{Data Analysis}

In this section, we deep dive into the ProCIS dataset to analyze its characteristics and understand its potential impact on the research community. 

\paragraph{\textbf{Utterance Length}} We analyze the distribution of comment lengths in the ProCIS dataset (Figure \ref{fig:comment-lengths-plot}). We tokenize the data using segtok tokenizer from the NLTK library,\footnote{\url{https://www.nltk.org}} and found that 95\% of the comments have up to 200 tokens in length. As presented in Table~\ref{dataset-stats}, the average comment length is $70.54 \pm 82.38$. This shows that the lengths of comments can vary significantly, and sometimes can be quite long. This insight is essential because it denotes that modeling conversations require long context modeling.  It also highlights the need for retrieval models to be able to handle varying lengths of comments in order to effectively address the needs of users. 

\paragraph{\textbf{Conversation Length}} Around 95\% of the conversations in the ProCIS dataset have up to 700 tokens in length and consist of 5.41 turns on average with the standard deviation of 7.81 (see Figures \ref{fig:thread-token-lengths-plot} and \ref{fig:turns-plot} and Table~\ref{dataset-stats}). However, a small percentage of conversations can span over thousands of tokens and sometimes more than 20 turns. This demonstrates the complexity and dynamics of real conversations. It emphasizes the need for retrieval models to be capable of maintaining context and coherence across multiple conversational turns while providing relevant information in a timely manner. The test set, by design, has shorter conversations, so that it is easier to get good relevance assessments from crowdworkers.

\begin{figure}[t]
  \centering
  \includegraphics[width=\linewidth]{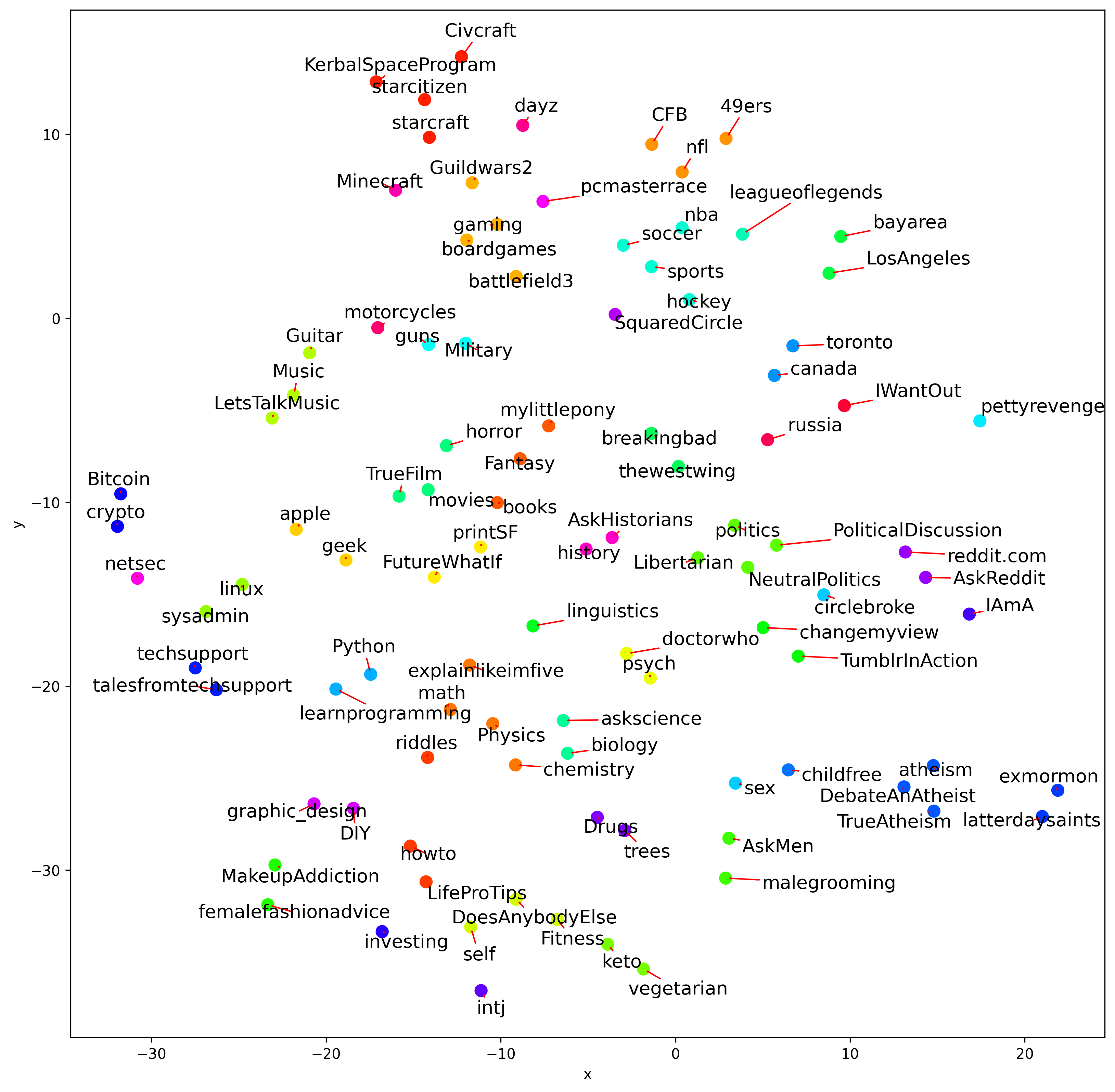}
  \caption{t-SNE visualization of the categories (subreddits) in the ProCIS test set.}
  \label{fig:points-plot}
\end{figure}

\paragraph{\textbf{Diversity of Topics}} ProCIS covers a broad range of topics, with the most popular categories being politics, religion, sports, finance, science, and general discussions (see Table~\ref{subreddits}). This extensive topical diversity reinforces the importance of developing retrieval models that can handle open-domain conversations. It also showcases the potential for ProCIS to serve as a benchmark for evaluating the performance of retrieval models across different domains. In Figure \ref{fig:points-plot}, we can see a clustered t-SNE \cite{hinton2002stochastic} projection of all the 100 subreddits covered in the test set. We use a pre-trained sentence embedding model\footnote{\url{https://hf.co/sentence-transformers/all-mpnet-base-v2}\label{sbert}} for encoding their descriptions and agglomerative clustering to identify broader groups. From the visualization we can see all the topic clusters covered in the test set. 

\section{Task Formulation}
In conversational information seeking systems, such as conversational question answering or conversational passage retrieval as modeled by the TREC CAsT Track \cite{owoicho2022trec}, the last user utterance is often a query or a question, and the goal is to retrieve a list of passages or produce an answer to the question in the last utterance in the context of the conversation. Orthogonal to these research problems, we target \emph{document retrieval as a form of contextual suggestion}. This means that the user utterances are not queries or questions and we do not aim at retrieving documents in order to answer them. Instead, we aim at retrieving documents that can add value to an ongoing conversation. Imagine there is a multi-party human conversation, and a retrieval engine can occasionally provide resources to contribute to the conversation. In this context, we identify two different retrieval tasks, as follows:
\begin{enumerate}
    \item \textbf{Reactive document retrieval for contextual suggestion in conversation:} During an ongoing multi-party human conversation, imagine any of the involved parties can hit a button to ask for recommendation or useful resources. These resources are explicit answers to any question but can contribute to the ongoing conversation. Since this document retrieval model is initiated by pushing a button, we call this a \emph{reactive} model.

    \item \textbf{Proactive document retrieval for contextual suggestion in conversation:} Imagine an agent is monitoring an ongoing multi-party human conversation and at any turn in the conversation can jump in and provide a useful suggestion by retrieving documents. These systems are more challenging than reactive systems, as they need to additionally decide when is a good time to proactively engage with the conversation.
\end{enumerate}

\paragraph{\textbf{Evaluating Reactive Document Retrieval Models}}
Given a conversation as a sequence of utterances $U = \{u_1, u_2, \cdots, u_m\}$ and a corpus of Wikipedia articles $D = \{d_1, d_2, \cdots, d_n\}$. The goal of reactive document retrieval  is to develop a retrieval model $M$ that takes a conversation up to turn $k$ (i.e., $\{u_1, u_2, \cdots, u_k\}$ and returns a ranked list of documents. We aim at providing useful suggestions that can contribute to any of the utterances in the given conversation. 

For evaluating reactive retrieval models, we assume the user asks for contextual suggestion at the end of the conversation, thus we use the whole conversation as query and retrieve documents. We then calculate standard retrieval metrics, such as nDCG@k, MRR, MAP, and Recall@k, to evaluate the system. Given the nature of conversational information seeking, precision-oriented metrics (with small $k$ values) are more important.

\begin{figure}[h!]
  \centering
  \includegraphics[width=0.47\textwidth]{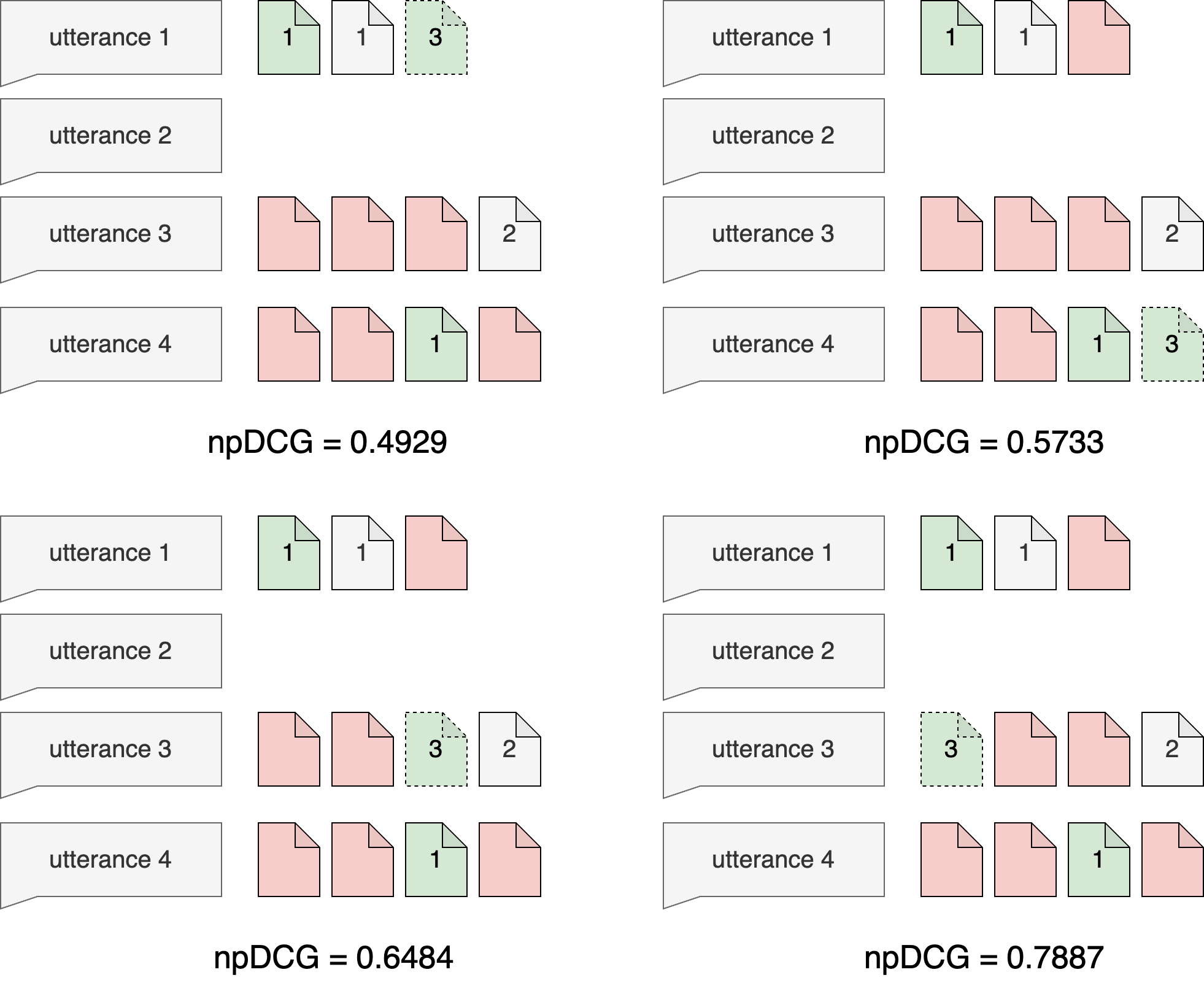}
  \caption{\small Examples of rank lists and their npDCG scores based on various positions of the doc with ideal position in utterance 3 (with dashed border). Green, gray and red represent relevant, partially relevant and irrelevant docs respectively.}
  \label{fig:npdcg}
\end{figure}

\paragraph{\textbf{Evaluating Proactive Document Retrieval Models}}
In proactive document retrieval, for each turn in a conversation, the goal is to either wait (do nothing) or retrieve a list of documents.
We evaluate proactive document retrieval as follows. At every conversation turn, a proactive retrieval system can make a binary decision: whether to retrieve and show a result list to the users or not. If the system decides to pass and not engage in the conversation, there is no cost to the users and also no gain is obtained. Therefore, we only evaluate the system when it presents retrieval results to the users. We can adopt different ranking metrics for this purpose, but in this paper, we only adopt nDCG as follows. Assume that a proactive retrieval model returns a result list $D_i = \{d_{i1}, d_{i2}, ..., d_{ik}\}$ after observing the $i$\textsuperscript{th} utterance in the conversation, i.e., $u_i$. Assume that $r_{ij} \in \{0, 1, 2\}$ denotes the relevance label associated with $d_{ij}$,
\begin{equation}
    pDCG = \frac{1}{Z} \sum_{i=1}^{n} \mathbbm{1}\{|D_i| > 0\} * DCG(D_i \setminus  \bigcup_{i'=1}^{i-1}{D_{i'}})
\end{equation}
where $\mathbbm{1}\{|D_i| > 0\}$ means if the proactive retrieval model engages in the conversation by retrieving a result list, i.e., if the result list is not empty. $Z$ is a normalization term and is equal to the number of times that the proactive retrieval model returns a result list, meaning $Z = \sum_{i=1}^{n} \mathbbm{1}\{|D_i| > 0\}$. The notation $D_i \setminus  \bigcup_{i'=1}^{i-1}{D_{i'}}$ means the result list returned in response to the $i$\textsuperscript{th} utterance excluding any result list that has been presented to the users before. The reason for this decision is that we assume there is no value in retrieving the same document over and over in the same dialogue and we do not want to reward a retrieval model that returns the same relevant document at multiple turns. Discounted Cumulative Gain (DCG) is calculated as:
\[
DCG(D_i) = \sum_{j=1}^{k} \frac{\text{rel}(r_{ij})}{\log(j+1)}
\]
where $r_{ij}$ is defined as follows:
\[
\text{rel}(r_{ij}) = \begin{cases} 
r_{ij} \times \frac{1}{\log(1+i-(l-1))} & \text{if } i \geq l \\
0 & \text{if } i < l
\end{cases}
\]
where $l$ is the perfect utterance number for document $d_{ij}$ to appear. In fact, if document $d_{ij}$ adds value to a conversation, turn $l$ is the first utterance in which this document becomes relevant. Therefore, if $i<l$ the document is non-relevant. Otherwise, if $i=l$, then, the model should not be penalized, thus $\text{rel}(r_{ij}) = r_{ij} \times \frac{1}{\log(2)} = r_{ij}$. If $i > l$, this means that the model has a delay in presenting a relevant document to its users and thus needs to be penalized. Inspired by the DCG formulation, we use logarithm as a concave penalization function. We next normalize $pDCG$ values as follows:
\begin{equation}
    npDCG = \frac{pDCG}{ipDCG}
\end{equation}
where $ipDCG$ denotes ideal $pDCG$ and represents the highest value that $pDCG$ can obtain for a given utterance. Therefore, $npDCG \in [0, 1]$. 
Note that $ipDCG$ is obtained by a model that for every turn retrieves all relevant documents, sorted by their relevance score, for that turn and if there is no relevant document associated with that turn, it skips retrieval. Figure~\ref{fig:npdcg} presents a toy example to explain how npDCG values are computed.

\begin{figure*}[t]
  \centering
  \includegraphics[width=0.77\textwidth]{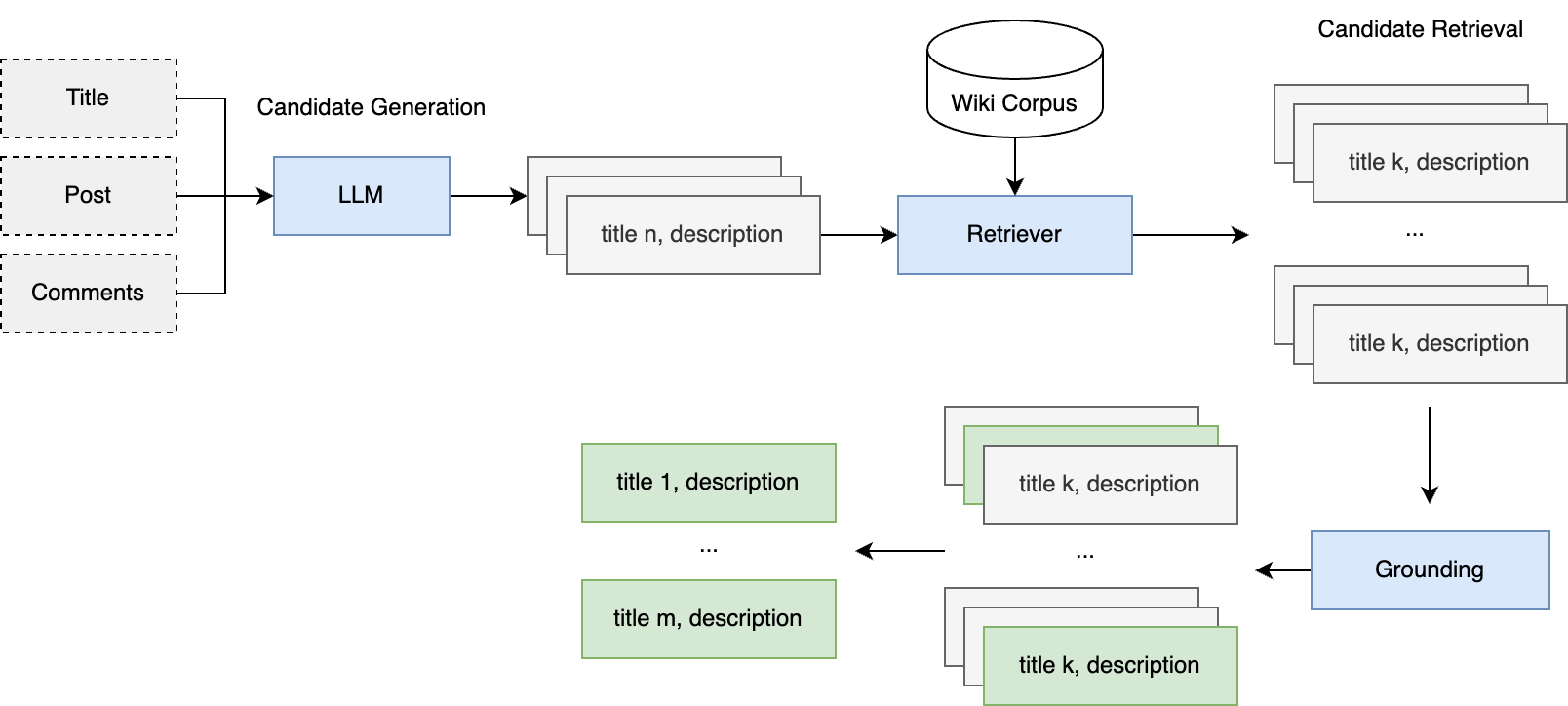}
  \caption{An overview of the Language Model Grounded Retrieval (LMGR) framework.}
  \label{fig:lmgr}
\end{figure*}

\section{Retrieval Methods}
\label{sec:method}
To provide benchmark results, we evaluate a wide range of methods on our datasets. Four of them are well-known existing methods from different categories, such as term-matching, neural sparse retrieval, neural single-vector dense retrieval, and neural multi-vector dense retrieval. We also introduce a novel approach specifically designed for this task that deals with a long sequence as input, called LMGR. 

\textbf{BM25} is a widely used ranking function that improves upon TF-IDF by incorporating term frequency saturation and document length normalization. It is the most common baseline in information retrieval tasks. For BM25 retrieval in our experiments, we used tantivy.\footnote{\url{https://github.com/quickwit-oss/tantivy}}

\textbf{SPLADE} is a neural retrieval model that uses sparse representations for documents and queries. It is based on explicit sparsity regularization and a log-saturation effect on term weights, leading to highly sparse representations. This model offers a trade-off between effectiveness and efficiency, and is trained end-to-end in a single stage \cite{ns4}. For SPLADE, as well as ColBERT below, we relied on the implementations from the tevatron library\cite{Gao2022TevatronAE} for our experiments.

\textbf{ANCE} is a dual-encoder neural retrieval model that uses dense representations to measure the similarity between queries and documents. It addresses the discrepancy between the data distribution used in training and testing by using an Approximate Nearest Neighbor (ANN) index of the corpus to select more realistic negative training instances \cite{ns2}. We used the official implementation\footnote{\url{https://github.com/microsoft/ANCE}} for training.

\textbf{ColBERT} is multi-vector dense retrieval model. It introduces a late interaction architecture that independently encodes the query and the document using BERT\cite{bert} and then employs a cheap yet powerful interaction step that models their similarity \cite{ns3}.

\medskip 

\textbf{Language Model Grounded Retrieval.}
The Language Model Grounded Retrieval (LMGR) framework, which we propose for this task, is inspired by the two-stage zero-shot entity linking approach of the BLINK model \cite{blink}. For each entity mentioned in a text, BLINK first retrieves some candidates using a dense retrieval model and then re-ranks them with a cross-encoder to pick the final result. Large Language Models have a good memorization of the Wikipedia corpus and can be very effective in generating relevant concepts to given complex queries. However, they can hallucinate and produce entities that are not actually in the Wikipedia corpus, or they can generate titles that are not exact matches. Applying a linking methodology like BLINK can solve this issue.

The LMGR framework consists of three stages: top-n candidate generation, top-k candidate retrieval, and grounding (Figure \ref{fig:lmgr}). 

\begin{itemize}[leftmargin=*]
    \item \textbf{Candidate Generation} In the first stage, the LMGR framework uses a large language model (LLM) to generate top-n candidates. The LLM is trained to predict the next word in a sentence, given the previous words. This allows it to generate a list of potential candidates using some separation format which could be relevant to the conversation. The LLM is capable of understanding the context of the conversation and generating candidates that are not only relevant but also diverse, addressing the complexity and diversity of open-domain conversations to a high extent. For each candidate, we prompt the LLM to generate pairs of title and one sentence descriptions. 

    \item \textbf{Candidate Retrieval} The second stage involves top-k candidate retrieval from each generated candidate from the LLM. This stage is crucial for identifying the corpus items that are closest to the generated candidate. For this stage, we employ dense retrieval using pre-trained sentence embeddings\textsuperscript{\ref{sbert}} from a corpus of Wikipedia title-description pairs, where description is limited to the first sentence of each Wikipedia article. 

    \item \textbf{Grounding} The third stage of the LMGR framework is the final candidate selection or grounding. In this stage, the LLM or a re-ranker is used to select the final candidate from the top-k candidates retrieved in the previous stage. The selected candidate is the one that the LLM or re-ranker determines to be the one most likely describing the same concept with the generated candidate.

\end{itemize}

In our experiments, we used the same LLM for both candidate generation and grounding. The LLM we used is based on the OpenChat-3.5 \cite{wang2023openchat}, which is a fine-tuned version of Mistral-7B \cite{jiang2023mistral}. We prompt the model to generate up to 20 candidates and experiment with retrieving and grounding 1, 3 and 5 results.

\subsection{Proactive Retrieval}
In addition to the retrieval methods mentioned above, we also employ a proactive retrieval approach that decides when to retrieve documents based on the current utterance and the conversation history. This is achieved using a binary classifier based on DeBERTa \cite{deberta} that predicts whether a given utterance requires document retrieval or not. The classifier is trained on pairs sampled in a balanced way from the training set. For each positive pair (an utterance with associated relevant documents), we randomly sample an utterance without associated relevant documents to create a negative pair. This balanced sampling helps the classifier learn to distinguish between utterances that require retrieval and those that do not.

The proactive retrieval classifier is applied before the actual retrieval methods. If the classifier predicts that an utterance requires retrieval, the selected retrieval method (e.g. ColBERT or LMGR) is then used to fetch relevant documents. This approach helps to reduce unnecessary retrieval for utterances that do not require additional information, thereby improving the efficiency and effectiveness of the overall system.

\section{Benchmark Results}

\paragraph{\textbf{Reactive Retrieval}}
Table~\ref{tab:results} shows the experimental results for reactive retrieval. BM25 is the worst performing model for this task and is expected, because term matching is far from sufficient to provide articles with useful context. Only for frequent entity mentions in a conversation this model might work to an extent. From the supervised neural models, ColBERT has the best performance, however it still struggles to achieve good recall metrics in the reactive setting. On the other hand, our proposed LMGR framework outperforms all the other models by a big margin in reactive retrieval, achieving very promising recall metrics as well with recall@5 up to 28.53\% and recall@20 53.06\%. This is an insightful finding, because this is one of the first retrieval tasks where zero-shot LLMs outperform existing state-of-the-art retrieval models by such a big margin. This showcases the potential of using LLMs directly for retrieval instead of using traditional techniques that involve vector representation and scoring. Generative retrieval models \cite{ns5,ns6} are already following this path.

\paragraph{\textbf{Proactive Retrieval}}
Table~\ref{tab:results_proactive} shows the experimental results for proactive retrieval. In contrast to the reactive retrieval setting, traditional retrieval models seem to outperform LMGR in proactive retrieval. This shows that LMGR is only suited for high-level understanding of a conversation and without fine-tuning can be unstable for proactive retrieval. The supervised neural models, particularly ColBERT, demonstrate better performance in the proactive setting compared to LMGR. This suggests that while LLMs excel in understanding the context and retrieving relevant articles reactively, they may struggle to anticipate future information needs without further adaptation to the specific task of proactive retrieval.


\begin{table*}[ht]
\caption{Experimental results for reactive retrieval on the ProCIS test set. The top section is for baselines and the bottom is our proposed LMGR framework. The superscript * denotes statistically significant improvements compared to all the baselines. $k$ is the number of retrieved candidates. Note that LMGR produces up to 20 results.}
\label{tab:results}
\begin{tabular}{lcccccccccc}
\toprule
\textbf{Model} & \textbf{nDCG@5} & \textbf{nDCG@20} & \textbf{nDCG@100} & \textbf{MRR} & \textbf{MAP} & \textbf{R@5} & \textbf{R@20} & \textbf{R@100} & \textbf{R@1K} \\
\midrule
BM25 & 0.0654 & 0.0754 & 0.0969 & 0.1561 & 0.0395 & 0.0410 & 0.0687 & 0.1202 & 0.2266 \\
SPLADE & 0.1605 & 0.1578 & 0.1575 & 0.4752 & 0.0752 & 0.0946 & 0.1343 & 0.1432 & 0.2946 \\
ANCE & 0.1854 & 0.1912 & 0.2240 & 0.4902 & 0.0984 & 0.0989 & 0.1635 & 0.2517 & 0.4316 \\
ColBERT & 0.2091 & 0.2094 & 0.2383 & 0.5679 & 0.1113 & 0.1117 & 0.1778 & 0.2649 & 0.4564 \\
\midrule
LMGR, k=1 & 0.2638 & 0.3678 & - & 0.6187 & 0.2000 & 0.2116 & 0.4091 & - & - \\
LMGR, k=3 & 0.2714 & 0.3986 & - & 0.6132 & 0.2198 & 0.2354 & 0.4614 & - & - \\
{LMGR, k=5} & \textbf{0.3408$^{*}$} & \textbf{0.4524$^{*}$} & - & \textbf{0.6300$^{*}$} & \textbf{0.2663$^{*}$} & \textbf{0.2853$^{*}$} & \textbf{0.5306$^{*}$} & - & - \\
\bottomrule
\end{tabular}
\vspace{0.2cm}
\centering
\end{table*}

\begin{table}[h]
\caption{Experimental results for proactive retrieval on the ProCIS test set using a DeBERTa-base proactive classifier. npDCG is the metric we defined for conversational proactive retrieval evaluation. The superscript * denotes statistically significant improvements compared to all the baselines. Note that LMGR produces up to 20 results.}
\label{tab:results_proactive}
\begin{tabular}{lcccc}
\toprule
\textbf{Model} & \textbf{npDCG@5} & \textbf{npDCG@20} & \textbf{npDCG@100} \\
\midrule
BM25 & 0.0229 & 0.0337 & 0.0405 \\
SPLADE & 0.1305 &	0.1440 & 0.1542 \\
ANCE & 0.1508 & 0.1792 & 0.2061 \\
{ColBERT} & \textbf{0.1719} & \textbf{0.1944} & \textbf{0.2172} \\ \midrule
LMGR, k=1 & 0.0574 & 0.1445 & - \\
LMGR, k=3 & 0.0613 & 0.1527 & - \\
LMGR, k=5 & 0.0781 & 0.1840 & - \\
\bottomrule
\end{tabular}
\centering
\end{table}

\section{Future Research Directions}

The ProCIS dataset serves as a foundation and starting point for future research in proactive CIS. However, there are several promising avenues for future work that can be explored. 

\textbf{Development of Proactive Retrieval Models} The unique nature of conversations, with their length, complexity, and domain diversity, presents a challenge for traditional retrieval models. Future research could focus on developing new models that are specifically designed to handle these challenges. For instance, the LMGR framework currently relies on large language models (LLMs), which can be computationally expensive. However, it's possible that smaller more cost-effective language models, trained on synthetic data, could perform just as well, if not better. Future research could also focus more on pro-activeness, exploring when to retrieve documents and how many to select.

\textbf{Improvement of Dense Retrieval Models} The potential for improvement in dense retrieval models is another promising area for future research. For example, a conversation encoder and a document encoder could be pre-trained separately and then fine-tuned on the target dataset. This could improve the model's ability to understand and respond to conversational context. Additionally, pre-training techniques for the conversation encoder, such as response ranking or response masking tasks, could be investigated. These techniques could further enhance the encoder's understanding of conversational dynamics.

\textbf{Advanced Pooling Methods} To capture the content of longer conversations more effectively, advanced pooling methods could be explored. Techniques such as averaging, attention mechanisms over utterance-level representations, and content filtering could be employed. These methods could help to distill the most important information from lengthy conversations, improving the model's ability to respond appropriately.

\textbf{Explainability} The utility of suggested concepts could be improved by generating explanations that clarify their relevance and importance within the context of the conversation. This could help users to better understand why certain concepts are being suggested, improving their overall experience.

\textbf{Query Generation} In addition to concept suggestions, the generation of natural language queries could be explored. This could further facilitate the information seeking process, making it easier for users to find the information they need.

\textbf{Synthetic Data Generation} The potential of LLMs for synthetic data generation is another area that could be investigated. Understanding how synthetic annotations can improve the performance of retrieval and generative models could be beneficial. This could lead to more accurate and efficient models.

\textbf{Generative Retrieval Models} Finally, experimenting with generative retrieval models \cite{ns5,ns6} for this task could also be a promising direction for future work. These models could potentially provide more accurate and relevant responses, improving the overall user experience.

\section{Conclusions}
In this paper, we have introduced ProCIS, a large-scale dataset for proactive conversational information seeking  collected from Reddit threads and enriched with external links to Wikipedia articles. ProCIS addresses a significant gap in the research landscape, providing a standardized benchmark for the development and evaluation of proactive retrieval models in the context of open-domain conversations. The dataset consists of over 2.8 million multi-party conversations, offering a rich resource for exploring the complexities and challenges of proactive IR in conversational settings.

We also proposed the Language Model Grounded Retrieval framework (LMGR) as a  baseline for this new task. Despite being a zero-shot method in our current experiments, LMGR outperforms existing ad-hoc retrieval models by a significant margin in the reactive setting, showing that after optimization this might be a very effective approach in the proactive setting as well. This also showcases the potential of using LLMs directly for retrieval instead of using traditional techniques that involve embedding and scoring.


The ProCIS dataset represents a significant step forward in the advancement of conversational agents capable of proactively seeking out and providing useful information to users. We hope that this dataset will inspire further research in the area of proactive conversational search and lead to the emergence of new techniques and approaches that will enhance user experiences in conversations and unlock the full potential of conversational LLM agents in various domains.

\section*{Acknowledgments}
This work was supported in part by the Center for Intelligent Information Retrieval, in part by the Office of Naval Research contract number N000142212688, in part by NSF grant number 2143434, in part by the Amazon Alexa Prize Competition, and in part by an award from Adobe.  Any opinions, findings and conclusions or recommendations expressed in this material are those of the authors and do not necessarily reflect those of the sponsor.

\newpage

\bibliographystyle{ACM-Reference-Format}
\bibliography{references}


\begin{thebibliography}{50}


\ifx \showCODEN    \undefined \def \showCODEN     #1{\unskip}     \fi
\ifx \showDOI      \undefined \def \showDOI       #1{#1}\fi
\ifx \showISBNx    \undefined \def \showISBNx     #1{\unskip}     \fi
\ifx \showISBNxiii \undefined \def \showISBNxiii  #1{\unskip}     \fi
\ifx \showISSN     \undefined \def \showISSN      #1{\unskip}     \fi
\ifx \showLCCN     \undefined \def \showLCCN      #1{\unskip}     \fi
\ifx \shownote     \undefined \def \shownote      #1{#1}          \fi
\ifx \showarticletitle \undefined \def \showarticletitle #1{#1}   \fi
\ifx \showURL      \undefined \def \showURL       {\relax}        \fi
\providecommand\bibfield[2]{#2}
\providecommand\bibinfo[2]{#2}
\providecommand\natexlab[1]{#1}
\providecommand\showeprint[2][]{arXiv:#2}

\bibitem[Ahmed and Bulathwela(2022)]%
        {rel5}
\bibfield{author}{\bibinfo{person}{Tabish Ahmed} {and} \bibinfo{person}{Sahan Bulathwela}.} \bibinfo{year}{2022}\natexlab{}.
\newblock \showarticletitle{Towards Proactive Information Retrieval in Noisy Text with Wikipedia Concepts}.
\newblock \bibinfo{journal}{\emph{arXiv preprint arXiv:2210.09877}} (\bibinfo{year}{2022}).
\newblock


\bibitem[Aliannejadi et~al\mbox{.}(2021)]%
        {10.1145/3459637.3482231}
\bibfield{author}{\bibinfo{person}{Mohammad Aliannejadi}, \bibinfo{person}{Leif Azzopardi}, \bibinfo{person}{Hamed Zamani}, \bibinfo{person}{Evangelos Kanoulas}, \bibinfo{person}{Paul Thomas}, {and} \bibinfo{person}{Nick Craswell}.} \bibinfo{year}{2021}\natexlab{}.
\newblock \showarticletitle{Analysing Mixed Initiatives and Search Strategies during Conversational Search}. In \bibinfo{booktitle}{\emph{Proceedings of the 30th ACM International Conference on Information \& Knowledge Management}} (Virtual Event, Queensland, Australia) \emph{(\bibinfo{series}{CIKM '21})}. \bibinfo{publisher}{Association for Computing Machinery}, \bibinfo{address}{New York, NY, USA}, \bibinfo{pages}{16–26}.
\newblock
\showISBNx{9781450384469}
\urldef\tempurl%
\url{https://doi.org/10.1145/3459637.3482231}
\showDOI{\tempurl}


\bibitem[Aliannejadi et~al\mbox{.}(2019)]%
        {Aliannejadi2019Qulac}
\bibfield{author}{\bibinfo{person}{Mohammad Aliannejadi}, \bibinfo{person}{Hamed Zamani}, \bibinfo{person}{Fabio Crestani}, {and} \bibinfo{person}{W.~Bruce Croft}.} \bibinfo{year}{2019}\natexlab{}.
\newblock \showarticletitle{Asking Clarifying Questions in Open-Domain Information-Seeking Conversations}. In \bibinfo{booktitle}{\emph{Proceedings of the 42nd International ACM SIGIR Conference on Research and Development in Information Retrieval}} (Paris, France) \emph{(\bibinfo{series}{SIGIR'19})}. \bibinfo{publisher}{Association for Computing Machinery}, \bibinfo{address}{New York, NY, USA}, \bibinfo{pages}{475–484}.
\newblock
\showISBNx{9781450361729}
\urldef\tempurl%
\url{https://doi.org/10.1145/3331184.3331265}
\showDOI{\tempurl}


\bibitem[Andolina et~al\mbox{.}(2018)]%
        {rel1}
\bibfield{author}{\bibinfo{person}{Salvatore Andolina}, \bibinfo{person}{Valeria Orso}, \bibinfo{person}{Hendrik Schneider}, \bibinfo{person}{Khalil Klouche}, \bibinfo{person}{Tuukka Ruotsalo}, \bibinfo{person}{Luciano Gamberini}, {and} \bibinfo{person}{Giulio Jacucci}.} \bibinfo{year}{2018}\natexlab{}.
\newblock \showarticletitle{Investigating Proactive Search Support in Conversations}. In \bibinfo{booktitle}{\emph{Proceedings of the 2018 Designing Interactive Systems Conference}} (Hong Kong, China) \emph{(\bibinfo{series}{DIS '18})}. \bibinfo{publisher}{Association for Computing Machinery}, \bibinfo{address}{New York, NY, USA}, \bibinfo{pages}{1295–1307}.
\newblock
\showISBNx{9781450351980}
\urldef\tempurl%
\url{https://doi.org/10.1145/3196709.3196734}
\showDOI{\tempurl}


\bibitem[Arampatzis(2001)]%
        {arampatzis2009threshold}
\bibfield{author}{\bibinfo{person}{Avi Arampatzis}.} \bibinfo{year}{2001}\natexlab{}.
\newblock \showarticletitle{Unbiased sd threshold optimization, initial query degradation, decay, and incrementality, for adaptive document filtering.}. In \bibinfo{booktitle}{\emph{TREC}}.
\newblock


\bibitem[Besan{\c{c}}on et~al\mbox{.}(2009)]%
        {clefinfile}
\bibfield{author}{\bibinfo{person}{Romaric Besan{\c{c}}on}, \bibinfo{person}{St{\'e}phane Chaudiron}, \bibinfo{person}{Djamel Mostefa}, \bibinfo{person}{Olivier Hamon}, \bibinfo{person}{Isma{\"i}l Timimi}, {and} \bibinfo{person}{Khalid Choukri}.} \bibinfo{year}{2009}\natexlab{}.
\newblock \showarticletitle{Overview of CLEF 2008 INFILE Pilot Track}. In \bibinfo{booktitle}{\emph{Evaluating Systems for Multilingual and Multimodal Information Access}}, \bibfield{editor}{\bibinfo{person}{Carol Peters}, \bibinfo{person}{Thomas Deselaers}, \bibinfo{person}{Nicola Ferro}, \bibinfo{person}{Julio Gonzalo}, \bibinfo{person}{Gareth J.~F. Jones}, \bibinfo{person}{Mikko Kurimo}, \bibinfo{person}{Thomas Mandl}, \bibinfo{person}{Anselmo Pe{\~{n}}as}, {and} \bibinfo{person}{Vivien Petras}} (Eds.). \bibinfo{publisher}{Springer Berlin Heidelberg}, \bibinfo{address}{Berlin, Heidelberg}, \bibinfo{pages}{939--946}.
\newblock
\showISBNx{978-3-642-04447-2}


\bibitem[Braslavski et~al\mbox{.}(2017)]%
        {Braslavski:2017}
\bibfield{author}{\bibinfo{person}{Pavel Braslavski}, \bibinfo{person}{Denis Savenkov}, \bibinfo{person}{Eugene Agichtein}, {and} \bibinfo{person}{Alina Dubatovka}.} \bibinfo{year}{2017}\natexlab{}.
\newblock \showarticletitle{What Do You Mean Exactly? Analyzing Clarification Questions in CQA}. In \bibinfo{booktitle}{\emph{Proceedings of the 2017 Conference on Conference Human Information Interaction and Retrieval}} (Oslo, Norway) \emph{(\bibinfo{series}{CHIIR '17})}. \bibinfo{publisher}{Association for Computing Machinery}, \bibinfo{address}{New York, NY, USA}, \bibinfo{pages}{345–348}.
\newblock
\showISBNx{9781450346771}
\urldef\tempurl%
\url{https://doi.org/10.1145/3020165.3022149}
\showDOI{\tempurl}


\bibitem[Choi et~al\mbox{.}(2018)]%
        {choi-etal-2018-quac}
\bibfield{author}{\bibinfo{person}{Eunsol Choi}, \bibinfo{person}{He He}, \bibinfo{person}{Mohit Iyyer}, \bibinfo{person}{Mark Yatskar}, \bibinfo{person}{Wen-tau Yih}, \bibinfo{person}{Yejin Choi}, \bibinfo{person}{Percy Liang}, {and} \bibinfo{person}{Luke Zettlemoyer}.} \bibinfo{year}{2018}\natexlab{}.
\newblock \showarticletitle{{Q}u{AC}: Question Answering in Context}. In \bibinfo{booktitle}{\emph{Proceedings of the 2018 Conference on Empirical Methods in Natural Language Processing}}. \bibinfo{publisher}{Association for Computational Linguistics}, \bibinfo{address}{Brussels, Belgium}, \bibinfo{pages}{2174--2184}.
\newblock
\urldef\tempurl%
\url{https://doi.org/10.18653/v1/D18-1241}
\showDOI{\tempurl}


\bibitem[Dalton et~al\mbox{.}(2019)]%
        {rel7}
\bibfield{author}{\bibinfo{person}{Jeffrey Dalton}, \bibinfo{person}{Chenyan Xiong}, {and} \bibinfo{person}{Jamie Callan}.} \bibinfo{year}{2019}\natexlab{}.
\newblock \showarticletitle{TREC CAsT 2019: The conversational assistance track overview}.
\newblock \bibinfo{journal}{\emph{TREC}} (\bibinfo{year}{2019}).
\newblock


\bibitem[Devlin et~al\mbox{.}(2019)]%
        {bert}
\bibfield{author}{\bibinfo{person}{Jacob Devlin}, \bibinfo{person}{Ming-Wei Chang}, \bibinfo{person}{Kenton Lee}, {and} \bibinfo{person}{Kristina Toutanova}.} \bibinfo{year}{2019}\natexlab{}.
\newblock \showarticletitle{{BERT}: Pre-training of Deep Bidirectional Transformers for Language Understanding}. In \bibinfo{booktitle}{\emph{Proceedings of the 2019 Conference of the North {A}merican Chapter of the Association for Computational Linguistics: Human Language Technologies, Volume 1 (Long and Short Papers)}}. \bibinfo{publisher}{Association for Computational Linguistics}, \bibinfo{address}{Minneapolis, Minnesota}, \bibinfo{pages}{4171--4186}.
\newblock
\urldef\tempurl%
\url{https://doi.org/10.18653/v1/N19-1423}
\showDOI{\tempurl}


\bibitem[Formal et~al\mbox{.}(2021)]%
        {ns4}
\bibfield{author}{\bibinfo{person}{Thibault Formal}, \bibinfo{person}{Benjamin Piwowarski}, {and} \bibinfo{person}{St\'{e}phane Clinchant}.} \bibinfo{year}{2021}\natexlab{}.
\newblock \showarticletitle{SPLADE: Sparse Lexical and Expansion Model for First Stage Ranking}. In \bibinfo{booktitle}{\emph{Proceedings of the 44th International ACM SIGIR Conference on Research and Development in Information Retrieval}} (Virtual Event, Canada) \emph{(\bibinfo{series}{SIGIR '21})}. \bibinfo{publisher}{Association for Computing Machinery}, \bibinfo{address}{New York, NY, USA}, \bibinfo{pages}{2288–2292}.
\newblock
\showISBNx{9781450380379}
\urldef\tempurl%
\url{https://doi.org/10.1145/3404835.3463098}
\showDOI{\tempurl}


\bibitem[Ganguly et~al\mbox{.}(2020)]%
        {DBLP:conf/fire/GangulyPVS20}
\bibfield{author}{\bibinfo{person}{Debasis Ganguly}, \bibinfo{person}{Dipasree Pal}, \bibinfo{person}{Manisha Verma}, {and} \bibinfo{person}{Procheta Sen}.} \bibinfo{year}{2020}\natexlab{}.
\newblock \showarticletitle{Overview of RCD-2020, the {FIRE-2020} track on Retrieval from Conversational Dialogues}. In \bibinfo{booktitle}{\emph{{FIRE} 2020: Forum for Information Retrieval Evaluation, Hyderabad, India, December 16-20, 2020}}, \bibfield{editor}{\bibinfo{person}{Prasenjit Majumder}, \bibinfo{person}{Mandar Mitra}, \bibinfo{person}{Surupendu Gangopadhyay}, {and} \bibinfo{person}{Parth Mehta}} (Eds.). \bibinfo{publisher}{{ACM}}, \bibinfo{pages}{33--36}.
\newblock
\urldef\tempurl%
\url{https://doi.org/10.1145/3441501.3441518}
\showDOI{\tempurl}


\bibitem[Gao et~al\mbox{.}(2022)]%
        {Gao2022TevatronAE}
\bibfield{author}{\bibinfo{person}{Luyu Gao}, \bibinfo{person}{Xueguang Ma}, \bibinfo{person}{Jimmy~J. Lin}, {and} \bibinfo{person}{Jamie Callan}.} \bibinfo{year}{2022}\natexlab{}.
\newblock \showarticletitle{Tevatron: An Efficient and Flexible Toolkit for Dense Retrieval}.
\newblock \bibinfo{journal}{\emph{ArXiv}}  \bibinfo{volume}{abs/2203.05765} (\bibinfo{year}{2022}).
\newblock


\bibitem[Gauch et~al\mbox{.}(2007)]%
        {gauch2007profile}
\bibfield{author}{\bibinfo{person}{Susan Gauch}, \bibinfo{person}{Mirco Speretta}, \bibinfo{person}{Aravind Chandramouli}, {and} \bibinfo{person}{Alessandro Micarelli}.} \bibinfo{year}{2007}\natexlab{}.
\newblock \bibinfo{booktitle}{\emph{User Profiles for Personalized Information Access}}.
\newblock \bibinfo{publisher}{Springer Berlin Heidelberg}, \bibinfo{address}{Berlin, Heidelberg}, \bibinfo{pages}{54--89}.
\newblock
\showISBNx{978-3-540-72079-9}
\urldef\tempurl%
\url{https://doi.org/10.1007/978-3-540-72079-9_2}
\showDOI{\tempurl}


\bibitem[Hattimare et~al\mbox{.}(2023)]%
        {Amherst2023}
\bibfield{author}{\bibinfo{person}{Amit Hattimare}, \bibinfo{person}{Arkin Dharawat}, \bibinfo{person}{Yelman Khan}, \bibinfo{person}{Yen-Chieh Lien}, \bibinfo{person}{Chris Samarinas}, \bibinfo{person}{George~Z. Wei}, \bibinfo{person}{Yulin Yang}, {and} \bibinfo{person}{Hamed Zamani}.} \bibinfo{year}{2023}\natexlab{}.
\newblock \showarticletitle{MarunaBot: Multi-Modal Augmentation for Large Language Models with Applications to Task-Oriented Dialogues}. In \bibinfo{booktitle}{\emph{Alexa Prize TaskBot Challenge 2 Proceedings}}.
\newblock
\urldef\tempurl%
\url{https://www.amazon.science/alexa-prize/proceedings/marunabot-v2-towards-end-to-end-multi-modal-task-oriented-dialogue-systems}
\showURL{%
\tempurl}


\bibitem[He et~al\mbox{.}(2023)]%
        {deberta}
\bibfield{author}{\bibinfo{person}{Pengcheng He}, \bibinfo{person}{Jianfeng Gao}, {and} \bibinfo{person}{Weizhu Chen}.} \bibinfo{year}{2023}\natexlab{}.
\newblock \showarticletitle{DeBERTaV3: Improving DeBERTa using ELECTRA-Style Pre-Training with Gradient-Disentangled Embedding Sharing}. In \bibinfo{booktitle}{\emph{The Eleventh International Conference on Learning Representations, {ICLR} 2023, Kigali, Rwanda, May 1-5, 2023}}. \bibinfo{publisher}{OpenReview.net}.
\newblock
\urldef\tempurl%
\url{https://openreview.net/pdf?id=sE7-XhLxHA}
\showURL{%
\tempurl}


\bibitem[Hinton and Roweis(2002)]%
        {hinton2002stochastic}
\bibfield{author}{\bibinfo{person}{Geoffrey~E Hinton} {and} \bibinfo{person}{Sam Roweis}.} \bibinfo{year}{2002}\natexlab{}.
\newblock \showarticletitle{Stochastic neighbor embedding}.
\newblock \bibinfo{journal}{\emph{Advances in neural information processing systems}}  \bibinfo{volume}{15} (\bibinfo{year}{2002}).
\newblock


\bibitem[Jiang et~al\mbox{.}(2023)]%
        {jiang2023mistral}
\bibfield{author}{\bibinfo{person}{Albert~Q. Jiang}, \bibinfo{person}{Alexandre Sablayrolles}, \bibinfo{person}{Arthur Mensch}, \bibinfo{person}{Chris Bamford}, \bibinfo{person}{Devendra~Singh Chaplot}, \bibinfo{person}{Diego de~las Casas}, \bibinfo{person}{Florian Bressand}, \bibinfo{person}{Gianna Lengyel}, \bibinfo{person}{Guillaume Lample}, \bibinfo{person}{Lucile Saulnier}, \bibinfo{person}{Lélio~Renard Lavaud}, \bibinfo{person}{Marie-Anne Lachaux}, \bibinfo{person}{Pierre Stock}, \bibinfo{person}{Teven~Le Scao}, \bibinfo{person}{Thibaut Lavril}, \bibinfo{person}{Thomas Wang}, \bibinfo{person}{Timothée Lacroix}, {and} \bibinfo{person}{William~El Sayed}.} \bibinfo{year}{2023}\natexlab{}.
\newblock \bibinfo{title}{Mistral 7B}.
\newblock
\newblock
\showeprint[arxiv]{2310.06825}~[cs.CL]


\bibitem[Khattab and Zaharia(2020)]%
        {ns3}
\bibfield{author}{\bibinfo{person}{Omar Khattab} {and} \bibinfo{person}{Matei Zaharia}.} \bibinfo{year}{2020}\natexlab{}.
\newblock \showarticletitle{ColBERT: Efficient and Effective Passage Search via Contextualized Late Interaction over BERT}. In \bibinfo{booktitle}{\emph{Proceedings of the 43rd International ACM SIGIR Conference on Research and Development in Information Retrieval}} (Virtual Event, China) \emph{(\bibinfo{series}{SIGIR '20})}. \bibinfo{publisher}{Association for Computing Machinery}, \bibinfo{address}{New York, NY, USA}, \bibinfo{pages}{39–48}.
\newblock
\showISBNx{9781450380164}
\urldef\tempurl%
\url{https://doi.org/10.1145/3397271.3401075}
\showDOI{\tempurl}


\bibitem[Kumar and Callan(2020)]%
        {related7}
\bibfield{author}{\bibinfo{person}{Vaibhav Kumar} {and} \bibinfo{person}{Jamie Callan}.} \bibinfo{year}{2020}\natexlab{}.
\newblock \showarticletitle{Making Information Seeking Easier: An Improved Pipeline for Conversational Search}. In \bibinfo{booktitle}{\emph{Findings of the Association for Computational Linguistics: EMNLP 2020}}, \bibfield{editor}{\bibinfo{person}{Trevor Cohn}, \bibinfo{person}{Yulan He}, {and} \bibinfo{person}{Yang Liu}} (Eds.). \bibinfo{publisher}{Association for Computational Linguistics}, \bibinfo{address}{Online}, \bibinfo{pages}{3971--3980}.
\newblock
\urldef\tempurl%
\url{https://doi.org/10.18653/v1/2020.findings-emnlp.354}
\showDOI{\tempurl}


\bibitem[Liao et~al\mbox{.}(2023)]%
        {10.1145/3539597.3572724}
\bibfield{author}{\bibinfo{person}{Lizi Liao}, \bibinfo{person}{Grace~Hui Yang}, {and} \bibinfo{person}{Chirag Shah}.} \bibinfo{year}{2023}\natexlab{}.
\newblock \showarticletitle{Proactive Conversational Agents}. In \bibinfo{booktitle}{\emph{Proceedings of the Sixteenth ACM International Conference on Web Search and Data Mining}} (Singapore, Singapore) \emph{(\bibinfo{series}{WSDM '23})}. \bibinfo{publisher}{Association for Computing Machinery}, \bibinfo{address}{New York, NY, USA}, \bibinfo{pages}{1244–1247}.
\newblock
\showISBNx{9781450394079}
\urldef\tempurl%
\url{https://doi.org/10.1145/3539597.3572724}
\showDOI{\tempurl}


\bibitem[Mo et~al\mbox{.}(2024)]%
        {related11}
\bibfield{author}{\bibinfo{person}{Fengran Mo}, \bibinfo{person}{Chen Qu}, \bibinfo{person}{Kelong Mao}, \bibinfo{person}{Tianyu Zhu}, \bibinfo{person}{Zhan Su}, \bibinfo{person}{Kaiyu Huang}, {and} \bibinfo{person}{Jian-Yun Nie}.} \bibinfo{year}{2024}\natexlab{}.
\newblock \bibinfo{title}{History-Aware Conversational Dense Retrieval}.
\newblock
\newblock
\showeprint[arxiv]{2401.16659}~[cs.IR]


\bibitem[Nguyen et~al\mbox{.}(2016)]%
        {nguyen2016ms}
\bibfield{author}{\bibinfo{person}{Tri Nguyen}, \bibinfo{person}{Mir Rosenberg}, \bibinfo{person}{Xia Song}, \bibinfo{person}{Jianfeng Gao}, \bibinfo{person}{Saurabh Tiwary}, \bibinfo{person}{Rangan Majumder}, {and} \bibinfo{person}{Li Deng}.} \bibinfo{year}{2016}\natexlab{}.
\newblock \showarticletitle{MS MARCO: A Human Generated MAchine Reading COmprehension Dataset}.
\newblock  (\bibinfo{date}{November} \bibinfo{year}{2016}).
\newblock
\urldef\tempurl%
\url{https://www.microsoft.com/en-us/research/publication/ms-marco-human-generated-machine-reading-comprehension-dataset/}
\showURL{%
\tempurl}


\bibitem[OpenAI(2023)]%
        {openai2023gpt4}
\bibfield{author}{\bibinfo{person}{OpenAI}.} \bibinfo{year}{2023}\natexlab{}.
\newblock \bibinfo{title}{GPT-4 Technical Report}.
\newblock
\newblock
\showeprint[arxiv]{2303.08774}~[cs.CL]


\bibitem[Owoicho et~al\mbox{.}(2022)]%
        {owoicho2022trec}
\bibfield{author}{\bibinfo{person}{Paul Owoicho}, \bibinfo{person}{Jeffrey Dalton}, \bibinfo{person}{Mohammad Aliannejadi}, \bibinfo{person}{Leif Azzopardi}, \bibinfo{person}{Johanne~R Trippas}, {and} \bibinfo{person}{Svitlana Vakulenko}.} \bibinfo{year}{2022}\natexlab{}.
\newblock \showarticletitle{TREC CAsT 2022: Going beyond user ask and system retrieve with initiative and response generation}.
\newblock \bibinfo{journal}{\emph{NIST Special Publication}} (\bibinfo{year}{2022}), \bibinfo{pages}{500--338}.
\newblock


\bibitem[Penha et~al\mbox{.}(2019)]%
        {related4}
\bibfield{author}{\bibinfo{person}{Gustavo Penha}, \bibinfo{person}{Alexandru Balan}, {and} \bibinfo{person}{Claudia Hauff}.} \bibinfo{year}{2019}\natexlab{}.
\newblock \showarticletitle{Introducing mantis: a novel multi-domain information seeking dialogues dataset}.
\newblock \bibinfo{journal}{\emph{arXiv preprint arXiv:1912.04639}} (\bibinfo{year}{2019}).
\newblock


\bibitem[Qu et~al\mbox{.}(2020)]%
        {related5}
\bibfield{author}{\bibinfo{person}{Chen Qu}, \bibinfo{person}{Liu Yang}, \bibinfo{person}{Cen Chen}, \bibinfo{person}{Minghui Qiu}, \bibinfo{person}{W.~Bruce Croft}, {and} \bibinfo{person}{Mohit Iyyer}.} \bibinfo{year}{2020}\natexlab{}.
\newblock \showarticletitle{Open-Retrieval Conversational Question Answering}. In \bibinfo{booktitle}{\emph{Proceedings of the 43rd International ACM SIGIR Conference on Research and Development in Information Retrieval}} (Virtual Event, China) \emph{(\bibinfo{series}{SIGIR '20})}. \bibinfo{publisher}{Association for Computing Machinery}, \bibinfo{address}{New York, NY, USA}, \bibinfo{pages}{539–548}.
\newblock
\showISBNx{9781450380164}
\urldef\tempurl%
\url{https://doi.org/10.1145/3397271.3401110}
\showDOI{\tempurl}


\bibitem[Radlinski and Craswell(2017)]%
        {related1}
\bibfield{author}{\bibinfo{person}{Filip Radlinski} {and} \bibinfo{person}{Nick Craswell}.} \bibinfo{year}{2017}\natexlab{}.
\newblock \showarticletitle{A Theoretical Framework for Conversational Search}. In \bibinfo{booktitle}{\emph{Proceedings of the 2017 Conference on Conference Human Information Interaction and Retrieval}} (Oslo, Norway) \emph{(\bibinfo{series}{CHIIR '17})}. \bibinfo{publisher}{Association for Computing Machinery}, \bibinfo{address}{New York, NY, USA}, \bibinfo{pages}{117–126}.
\newblock
\showISBNx{9781450346771}
\urldef\tempurl%
\url{https://doi.org/10.1145/3020165.3020183}
\showDOI{\tempurl}


\bibitem[Rao and Daum{\'e}~III(2018)]%
        {Rao:2018}
\bibfield{author}{\bibinfo{person}{Sudha Rao} {and} \bibinfo{person}{Hal Daum{\'e}~III}.} \bibinfo{year}{2018}\natexlab{}.
\newblock \showarticletitle{Learning to Ask Good Questions: Ranking Clarification Questions using Neural Expected Value of Perfect Information}. In \bibinfo{booktitle}{\emph{Proceedings of the 56th Annual Meeting of the Association for Computational Linguistics (Volume 1: Long Papers)}}. \bibinfo{publisher}{Association for Computational Linguistics}, \bibinfo{address}{Melbourne, Australia}, \bibinfo{pages}{2737--2746}.
\newblock
\urldef\tempurl%
\url{https://doi.org/10.18653/v1/P18-1255}
\showDOI{\tempurl}


\bibitem[Reddy et~al\mbox{.}(2019)]%
        {reddy-etal-2019-coqa}
\bibfield{author}{\bibinfo{person}{Siva Reddy}, \bibinfo{person}{Danqi Chen}, {and} \bibinfo{person}{Christopher~D. Manning}.} \bibinfo{year}{2019}\natexlab{}.
\newblock \showarticletitle{{C}o{QA}: A Conversational Question Answering Challenge}.
\newblock \bibinfo{journal}{\emph{Transactions of the Association for Computational Linguistics}}  \bibinfo{volume}{7} (\bibinfo{year}{2019}), \bibinfo{pages}{249--266}.
\newblock
\urldef\tempurl%
\url{https://doi.org/10.1162/tacl_a_00266}
\showDOI{\tempurl}


\bibitem[Robertson and Soboroff(2003)]%
        {trecfiltering}
\bibfield{author}{\bibinfo{person}{S Robertson} {and} \bibinfo{person}{Ian Soboroff}.} \bibinfo{year}{2003}\natexlab{}.
\newblock \bibinfo{title}{The TREC-2002 Filtering Track Report}.
\newblock
\newblock
\urldef\tempurl%
\url{https://tsapps.nist.gov/publication/get_pdf.cfm?pub_id=50769}
\showURL{%
\tempurl}


\bibitem[Ros et~al\mbox{.}(2023)]%
        {10.1145/3578337.3605139}
\bibfield{author}{\bibinfo{person}{Kevin Ros}, \bibinfo{person}{Matthew Jin}, \bibinfo{person}{Jacob Levine}, {and} \bibinfo{person}{ChengXiang Zhai}.} \bibinfo{year}{2023}\natexlab{}.
\newblock \showarticletitle{Retrieving Webpages Using Online Discussions}. In \bibinfo{booktitle}{\emph{Proceedings of the 2023 ACM SIGIR International Conference on Theory of Information Retrieval}} (Taipei, Taiwan) \emph{(\bibinfo{series}{ICTIR '23})}. \bibinfo{publisher}{Association for Computing Machinery}, \bibinfo{address}{New York, NY, USA}, \bibinfo{pages}{159–168}.
\newblock
\showISBNx{9798400700736}
\urldef\tempurl%
\url{https://doi.org/10.1145/3578337.3605139}
\showDOI{\tempurl}


\bibitem[Samarinas et~al\mbox{.}(2024)]%
        {samarinas2024simulating}
\bibfield{author}{\bibinfo{person}{Chris Samarinas}, \bibinfo{person}{Pracha Promthaw}, \bibinfo{person}{Atharva Nijasure}, \bibinfo{person}{Hansi Zeng}, \bibinfo{person}{Julian Killingback}, {and} \bibinfo{person}{Hamed Zamani}.} \bibinfo{year}{2024}\natexlab{}.
\newblock \bibinfo{title}{Simulating Task-Oriented Dialogues with State Transition Graphs and Large Language Models}.
\newblock
\newblock
\showeprint[arxiv]{2404.14772}~[cs.CL]


\bibitem[Sekuli\'{c} et~al\mbox{.}(2022)]%
        {10.1145/3488560.3498440}
\bibfield{author}{\bibinfo{person}{Ivan Sekuli\'{c}}, \bibinfo{person}{Mohammad Aliannejadi}, {and} \bibinfo{person}{Fabio Crestani}.} \bibinfo{year}{2022}\natexlab{}.
\newblock \showarticletitle{Evaluating Mixed-initiative Conversational Search Systems via User Simulation}. In \bibinfo{booktitle}{\emph{Proceedings of the Fifteenth ACM International Conference on Web Search and Data Mining}} (Virtual Event, AZ, USA) \emph{(\bibinfo{series}{WSDM '22})}. \bibinfo{publisher}{Association for Computing Machinery}, \bibinfo{address}{New York, NY, USA}, \bibinfo{pages}{888–896}.
\newblock
\showISBNx{9781450391320}
\urldef\tempurl%
\url{https://doi.org/10.1145/3488560.3498440}
\showDOI{\tempurl}


\bibitem[Sen et~al\mbox{.}(2018)]%
        {rel3}
\bibfield{author}{\bibinfo{person}{Procheta Sen}, \bibinfo{person}{Debasis Ganguly}, {and} \bibinfo{person}{Gareth Jones}.} \bibinfo{year}{2018}\natexlab{}.
\newblock \showarticletitle{Procrastination is the Thief of Time: Evaluating the Effectiveness of Proactive Search Systems}. In \bibinfo{booktitle}{\emph{The 41st International ACM SIGIR Conference on Research \& Development in Information Retrieval}} (Ann Arbor, MI, USA) \emph{(\bibinfo{series}{SIGIR '18})}. \bibinfo{publisher}{Association for Computing Machinery}, \bibinfo{address}{New York, NY, USA}, \bibinfo{pages}{1157–1160}.
\newblock
\showISBNx{9781450356572}
\urldef\tempurl%
\url{https://doi.org/10.1145/3209978.3210114}
\showDOI{\tempurl}


\bibitem[Shah(2018)]%
        {rel4}
\bibfield{author}{\bibinfo{person}{Chirag Shah}.} \bibinfo{year}{2018}\natexlab{}.
\newblock \showarticletitle{Information fostering-being proactive with information seeking and retrieval: Perspective paper}. In \bibinfo{booktitle}{\emph{Proceedings of the 2018 conference on human information interaction \& retrieval}}. \bibinfo{pages}{62--71}.
\newblock


\bibitem[Stoyanchev et~al\mbox{.}(2014)]%
        {Stoyanchev:2014}
\bibfield{author}{\bibinfo{person}{Svetlana Stoyanchev}, \bibinfo{person}{Alex Liu}, {and} \bibinfo{person}{Julia Hirschberg}.} \bibinfo{year}{2014}\natexlab{}.
\newblock \showarticletitle{Towards Natural Clarification Questions in Dialogue Systems}. In \bibinfo{booktitle}{\emph{AISB '14}}, Vol.~\bibinfo{volume}{20}.
\newblock


\bibitem[Tay et~al\mbox{.}(2022)]%
        {ns5}
\bibfield{author}{\bibinfo{person}{Yi Tay}, \bibinfo{person}{Vinh Tran}, \bibinfo{person}{Mostafa Dehghani}, \bibinfo{person}{Jianmo Ni}, \bibinfo{person}{Dara Bahri}, \bibinfo{person}{Harsh Mehta}, \bibinfo{person}{Zhen Qin}, \bibinfo{person}{Kai Hui}, \bibinfo{person}{Zhe Zhao}, \bibinfo{person}{Jai Gupta}, \bibinfo{person}{Tal Schuster}, \bibinfo{person}{William~W Cohen}, {and} \bibinfo{person}{Donald Metzler}.} \bibinfo{year}{2022}\natexlab{}.
\newblock \showarticletitle{Transformer Memory as a Differentiable Search Index}. In \bibinfo{booktitle}{\emph{Advances in Neural Information Processing Systems}}, Vol.~\bibinfo{volume}{35}. \bibinfo{publisher}{Curran Associates, Inc.}, \bibinfo{pages}{21831--21843}.
\newblock
\urldef\tempurl%
\url{https://proceedings.neurips.cc/paper_files/paper/2022/file/892840a6123b5ec99ebaab8be1530fba-Paper-Conference.pdf}
\showURL{%
\tempurl}


\bibitem[Touvron et~al\mbox{.}(2023)]%
        {touvron2023llama}
\bibfield{author}{\bibinfo{person}{Hugo Touvron}, \bibinfo{person}{Thibaut Lavril}, \bibinfo{person}{Gautier Izacard}, \bibinfo{person}{Xavier Martinet}, \bibinfo{person}{Marie-Anne Lachaux}, \bibinfo{person}{Timothée Lacroix}, \bibinfo{person}{Baptiste Rozière}, \bibinfo{person}{Naman Goyal}, \bibinfo{person}{Eric Hambro}, \bibinfo{person}{Faisal Azhar}, \bibinfo{person}{Aurelien Rodriguez}, \bibinfo{person}{Armand Joulin}, \bibinfo{person}{Edouard Grave}, {and} \bibinfo{person}{Guillaume Lample}.} \bibinfo{year}{2023}\natexlab{}.
\newblock \bibinfo{title}{LLaMA: Open and Efficient Foundation Language Models}.
\newblock
\newblock
\showeprint[arxiv]{2302.13971}~[cs.CL]


\bibitem[Wadhwa and Zamani(2021)]%
        {related13}
\bibfield{author}{\bibinfo{person}{Somin Wadhwa} {and} \bibinfo{person}{Hamed Zamani}.} \bibinfo{year}{2021}\natexlab{}.
\newblock \showarticletitle{Towards System-Initiative Conversational Information Seeking.}. In \bibinfo{booktitle}{\emph{DESIRES}}. \bibinfo{pages}{102--116}.
\newblock


\bibitem[Wang et~al\mbox{.}(2023)]%
        {wang2023openchat}
\bibfield{author}{\bibinfo{person}{Guan Wang}, \bibinfo{person}{Sijie Cheng}, \bibinfo{person}{Xianyuan Zhan}, \bibinfo{person}{Xiangang Li}, \bibinfo{person}{Sen Song}, {and} \bibinfo{person}{Yang Liu}.} \bibinfo{year}{2023}\natexlab{}.
\newblock \showarticletitle{Openchat: Advancing open-source language models with mixed-quality data}.
\newblock \bibinfo{journal}{\emph{arXiv preprint arXiv:2309.11235}} (\bibinfo{year}{2023}).
\newblock


\bibitem[Wu et~al\mbox{.}(2020)]%
        {blink}
\bibfield{author}{\bibinfo{person}{Ledell Wu}, \bibinfo{person}{Fabio Petroni}, \bibinfo{person}{Martin Josifoski}, \bibinfo{person}{Sebastian Riedel}, {and} \bibinfo{person}{Luke Zettlemoyer}.} \bibinfo{year}{2020}\natexlab{}.
\newblock \showarticletitle{Scalable Zero-shot Entity Linking with Dense Entity Retrieval}. In \bibinfo{booktitle}{\emph{Proceedings of the 2020 Conference on Empirical Methods in Natural Language Processing (EMNLP)}}, \bibfield{editor}{\bibinfo{person}{Bonnie Webber}, \bibinfo{person}{Trevor Cohn}, \bibinfo{person}{Yulan He}, {and} \bibinfo{person}{Yang Liu}} (Eds.). \bibinfo{publisher}{Association for Computational Linguistics}, \bibinfo{address}{Online}, \bibinfo{pages}{6397--6407}.
\newblock
\urldef\tempurl%
\url{https://doi.org/10.18653/v1/2020.emnlp-main.519}
\showDOI{\tempurl}


\bibitem[Xiong et~al\mbox{.}(2021)]%
        {ns2}
\bibfield{author}{\bibinfo{person}{Lee Xiong}, \bibinfo{person}{Chenyan Xiong}, \bibinfo{person}{Ye Li}, \bibinfo{person}{Kwok{-}Fung Tang}, \bibinfo{person}{Jialin Liu}, \bibinfo{person}{Paul~N. Bennett}, \bibinfo{person}{Junaid Ahmed}, {and} \bibinfo{person}{Arnold Overwijk}.} \bibinfo{year}{2021}\natexlab{}.
\newblock \showarticletitle{Approximate Nearest Neighbor Negative Contrastive Learning for Dense Text Retrieval}. In \bibinfo{booktitle}{\emph{9th International Conference on Learning Representations, {ICLR} 2021, Virtual Event, Austria, May 3-7, 2021}}. \bibinfo{publisher}{OpenReview.net}.
\newblock
\urldef\tempurl%
\url{https://openreview.net/forum?id=zeFrfgyZln}
\showURL{%
\tempurl}


\bibitem[Yu et~al\mbox{.}(2021)]%
        {related3}
\bibfield{author}{\bibinfo{person}{Shi Yu}, \bibinfo{person}{Zhenghao Liu}, \bibinfo{person}{Chenyan Xiong}, \bibinfo{person}{Tao Feng}, {and} \bibinfo{person}{Zhiyuan Liu}.} \bibinfo{year}{2021}\natexlab{}.
\newblock \showarticletitle{Few-Shot Conversational Dense Retrieval}. In \bibinfo{booktitle}{\emph{Proceedings of the 44th International ACM SIGIR Conference on Research and Development in Information Retrieval}} (Virtual Event, Canada) \emph{(\bibinfo{series}{SIGIR '21})}. \bibinfo{publisher}{Association for Computing Machinery}, \bibinfo{address}{New York, NY, USA}, \bibinfo{pages}{829–838}.
\newblock
\showISBNx{9781450380379}
\urldef\tempurl%
\url{https://doi.org/10.1145/3404835.3462856}
\showDOI{\tempurl}


\bibitem[Zamani et~al\mbox{.}(2020a)]%
        {Zamani2020ClarGeneration}
\bibfield{author}{\bibinfo{person}{Hamed Zamani}, \bibinfo{person}{Susan Dumais}, \bibinfo{person}{Nick Craswell}, \bibinfo{person}{Paul Bennett}, {and} \bibinfo{person}{Gord Lueck}.} \bibinfo{year}{2020}\natexlab{a}.
\newblock \showarticletitle{Generating Clarifying Questions for Information Retrieval}. In \bibinfo{booktitle}{\emph{Proceedings of The Web Conference 2020}} (Taipei, Taiwan) \emph{(\bibinfo{series}{WWW '20})}. \bibinfo{publisher}{Association for Computing Machinery}, \bibinfo{address}{New York, NY, USA}, \bibinfo{pages}{418–428}.
\newblock
\showISBNx{9781450370233}
\urldef\tempurl%
\url{https://doi.org/10.1145/3366423.3380126}
\showDOI{\tempurl}


\bibitem[Zamani et~al\mbox{.}(2020b)]%
        {Zamani2020MIMICS}
\bibfield{author}{\bibinfo{person}{Hamed Zamani}, \bibinfo{person}{Gord Lueck}, \bibinfo{person}{Everest Chen}, \bibinfo{person}{Rodolfo Quispe}, \bibinfo{person}{Flint Luu}, {and} \bibinfo{person}{Nick Craswell}.} \bibinfo{year}{2020}\natexlab{b}.
\newblock \showarticletitle{MIMICS: A Large-Scale Data Collection for Search Clarification}. In \bibinfo{booktitle}{\emph{Proceedings of the 29th ACM International Conference on Information \& Knowledge Management}}. \bibinfo{publisher}{Association for Computing Machinery}, \bibinfo{address}{New York, NY, USA}, \bibinfo{pages}{3189–3196}.
\newblock
\showISBNx{9781450368599}
\urldef\tempurl%
\url{https://doi.org/10.1145/3340531.3412772}
\showURL{%
\tempurl}


\bibitem[Zamani et~al\mbox{.}(2020c)]%
        {Zamani2020ClarAnalysis}
\bibfield{author}{\bibinfo{person}{Hamed Zamani}, \bibinfo{person}{Bhaskar Mitra}, \bibinfo{person}{Everest Chen}, \bibinfo{person}{Gord Lueck}, \bibinfo{person}{Fernando Diaz}, \bibinfo{person}{Paul~N. Bennett}, \bibinfo{person}{Nick Craswell}, {and} \bibinfo{person}{Susan~T. Dumais}.} \bibinfo{year}{2020}\natexlab{c}.
\newblock \showarticletitle{Analyzing and Learning from User Interactions for Search Clarification}. In \bibinfo{booktitle}{\emph{Proceedings of the 43rd International ACM SIGIR Conference on Research and Development in Information Retrieval}} (Virtual Event, China) \emph{(\bibinfo{series}{SIGIR '20})}. \bibinfo{publisher}{Association for Computing Machinery}, \bibinfo{address}{New York, NY, USA}, \bibinfo{pages}{1181–1190}.
\newblock
\showISBNx{9781450380164}
\urldef\tempurl%
\url{https://doi.org/10.1145/3397271.3401160}
\showDOI{\tempurl}


\bibitem[Zamani and Shakery(2018)]%
        {contentbasedrecsys}
\bibfield{author}{\bibinfo{person}{Hamed Zamani} {and} \bibinfo{person}{Azadeh Shakery}.} \bibinfo{year}{2018}\natexlab{}.
\newblock \showarticletitle{A language model-based framework for multi-publisher content-based recommender systems}.
\newblock \bibinfo{journal}{\emph{Information Retrieval Journal}}  \bibinfo{volume}{21} (\bibinfo{year}{2018}), \bibinfo{pages}{369--409}.
\newblock
\urldef\tempurl%
\url{https://doi.org/10.1007/s10791-018-9327-0}
\showDOI{\tempurl}


\bibitem[Zamani et~al\mbox{.}(2023)]%
        {zamani2023conversational}
\bibfield{author}{\bibinfo{person}{Hamed Zamani}, \bibinfo{person}{Johanne~R. Trippas}, \bibinfo{person}{Jeff Dalton}, {and} \bibinfo{person}{Filip Radlinski}.} \bibinfo{year}{2023}\natexlab{}.
\newblock \showarticletitle{Conversational Information Seeking}.
\newblock \bibinfo{journal}{\emph{Foundations and Trends® in Information Retrieval}} \bibinfo{volume}{17}, \bibinfo{number}{3-4} (\bibinfo{year}{2023}), \bibinfo{pages}{244--456}.
\newblock
\showISSN{1554-0669}
\urldef\tempurl%
\url{https://doi.org/10.1561/1500000081}
\showDOI{\tempurl}


\bibitem[Zeng et~al\mbox{.}(2024)]%
        {ns6}
\bibfield{author}{\bibinfo{person}{Hansi Zeng}, \bibinfo{person}{Chen Luo}, \bibinfo{person}{Bowen Jin}, \bibinfo{person}{Sheikh~Muhammad Sarwar}, \bibinfo{person}{Tianxin Wei}, {and} \bibinfo{person}{Hamed Zamani}.} \bibinfo{year}{2024}\natexlab{}.
\newblock \showarticletitle{Scalable and Effective Generative Information Retrieval}. In \bibinfo{booktitle}{\emph{Proceedings of The Web Conference 2024}} (Singapore, Singapore) \emph{(\bibinfo{series}{WWW '24})}. \bibinfo{publisher}{Association for Computing Machinery}, \bibinfo{address}{New York, NY, USA}.
\newblock


\end{thebibliography}


\end{document}